\documentclass[a4paper,12pt,justification=centering]{article}
\usepackage{latexsym}
\usepackage{amsmath,amssymb,amsxtra,amscd,amsthm}
\usepackage[mathscr]{eucal}
\usepackage{graphicx}
\usepackage[justification=centering]{caption}
\usepackage{subfig}
\usepackage{datetime}
\usepackage{pdfsync}
\usepackage{thumbpdf}
\usepackage{color}
\ifx\pdfoutput\undefined
\usepackage
[dvips,
 verbose=false,
 bookmarks=true,
 colorlinks=true,
 linkcolor=webred,
 filecolor=webbrown,
 citecolor=webgreen,
 pagecolor=webblue,
 urlcolor=webblue,
 pdftitle={},
 pdfauthor={Murad Alim},
 pdfsubject={},
 pdfkeywords={},
 bookmarksopen=false,
 pdfpagemode=None,
 pdfview=FitH,
 pdfstartview=FitH,
 extension=pdf]{hyperref}
\else
\usepackage
[pdftex,
 verbose=false,
 bookmarks=true,
 colorlinks=true,
 linkcolor=webred,
 filecolor=webbrown,
 citecolor=webgreen,
 pagecolor=webblue,
 urlcolor=webblue,
 pdftitle={},
 pdfauthor={Emanuel Scheidegger},
 pdfsubject={},
 pdfkeywords={},
 bookmarksopen=false,
 pdfpagemode=None,
 pdfview=FitH,
 pdfstartview=FitH,
 extension=pdf]{hyperref}
\fi
\definecolor{webred}{rgb}{.8,0,0}
\definecolor{webbrown}{rgb}{.6,0,0}
\definecolor{webgreen}{rgb}{0,0.5,0}
\definecolor{webdkgreen}{rgb}{0,0.3,0}
\definecolor{webblue}{rgb}{0,0,0.5}

\numberwithin{equation}{section}

\providecommand{\href}[2]{#2}

\def\cE{\mathcal{E}}
\def\cF{\mathcal{F}}
\def\cL{\mathcal{L}}
\def\cM{\mathcal{M}}

\def\mC{\mathbb{C}}

\def\mZ{\mathbb{Z}}

\newcommand{\mE}{\mathcal{E}}

\newcommand{\F}[1]{\cF^{(#1)}}

\allowdisplaybreaks
\usepackage{bbm}
\usepackage{rotating}
\newcommand{\be}{\begin{eqnarray}}
\newcommand{\beq}{\begin{eqnarray}}
\newcommand{\ee}{\end{eqnarray}}

\newcommand{\h}{\frac{1}{2}}

\begin{document}

\setlength{\parindent}{0cm}
\setlength{\baselineskip}{1.5em}
\title{Special Polynomial Rings, Quasi Modular Forms\\ and Duality of Topological Strings}
\author{Murad Alim$^{1,2}$\footnote{\tt{alim@physics.harvard.edu}}, Emanuel Scheidegger$^3$\footnote{\tt{emanuel.scheidegger@math.uni-freiburg.de}}, Shing-Tung Yau$^1$\footnote{\tt yau@math.harvard.edu} and Jie Zhou$^1$\footnote{\tt jiezhou@math.harvard.edu}
\\
\small $^1$ Department of Mathematics, Harvard University,\\ \small 1 Oxford Street, Cambridge, MA 02138, USA\\
\small $^2$ Jefferson Physical Laboratory, Harvard University, \\ \small 17 Oxford Street, Cambridge, MA 02138, USA\\
\small $^3$ Mathematisches Institut, Albert-Ludwigs-Universit\"at Freiburg,\\ \small Eckerstrasse 1, D-79104 Freiburg, Germany }

\date{}
\maketitle

\abstract{We study the differential polynomial rings which are defined
  using the special geometry of the moduli spaces of Calabi-Yau
  threefolds. The higher genus topological string amplitudes are
  expressed as polynomials in the generators of these rings, giving them a global
  description in the moduli space. At particular loci, the amplitudes
  yield the generating functions of Gromov-Witten invariants. We show that these rings are isomorphic to the rings of quasi modular forms for threefolds with duality groups for which these are known. For
  the other cases, they provide generalizations thereof. We
  furthermore study an involution which acts on the
  quasi modular forms. We interpret it as a duality
  which exchanges two distinguished expansion loci of the topological string amplitudes in the moduli
  space. We construct these special polynomial rings and match them
  with known quasi modular forms for non-compact Calabi-Yau geometries
  and their mirrors including local $\mathbbm{P}^2$ and local del Pezzo geometries with
  $E_5,E_6,E_7$ and $E_8$ type singularities. We provide the analogous
  special polynomial ring for the quintic. }

\clearpage


\tableofcontents


\section{Introduction}

The study of physical theories within their moduli spaces has proven
to be a rich source of insights for mathematics and physics. For
$SU(2)$ gauge theory, the exact low energy effective energy has been
deduced in Ref.~\cite{Seiberg:1994rs} by understanding the
singularities in the moduli space due to a monopole and a dyon
becoming massless. Tracking the physical behavior of the theory has been achieved using the periods of an elliptic curve, whose monodromies take account of the various one loop contributions in different patches of the vacuum manifold of the theory or the moduli space. A key insight was the use of two different dual local coordinates on the moduli space $a,a_D$, depending on which region of moduli space is described. The physical coupling $\tau$ is computed as $\tau=\partial a_D/\partial a$ in the weak coupling, electric phase of the theory while the dual coupling $\tau_D$ in the magnetic phase is computed as $\tau_D=\partial a/\partial a_D$. The exchange of the two different expansion loci in the moduli as well as the two different theories attached to them is the $\mathcal{N}=2$ version of electric-magnetic duality.

The broader context to answer these types of questions about moduli
spaces of physical theories and their dualities is string theory. The
elliptic curve parameterizing the Seiberg-Witten solution can be
understood as part of a Calabi-Yau (CY) threefold
\cite{Klemm:1996bj}. The moduli space in the geometric context is that
of complex structures. The singular loci which are the analogs of the
loci where the monopoles became massless correspond to conifold
singularities where a three cycle shrinks to zero size
\cite{Candelas:1989js,Candelas:1990rm}. The physics of this singularity was understood in Ref.~\cite{Strominger:1995cz}.

Topological string theory provides a set of ideas and tools to study questions related to moduli spaces of theories, it allows one furthermore to use the power of mirror symmetry which identifies two deformation families of topological strings.\footnote{See Ref.~\cite{Alim:2012gq} and references therein.} The topological string partition function is defined as a perturbative sum over free energies associated to worldsheets of genus $g$ \cite{Bershadsky:1993cx}:
\begin{equation}\label{ztop}
Z_{top}(t,\bar{t})=\exp\left( \sum_{g=0}^{\infty} \lambda^{2g-2} \F{g}(t,\bar{t})\right)\,,
\end{equation}
where $\lambda$ is the topological string coupling constant and where the dependence on $t,\bar{t}$ stands for the dependence on a set of local coordinates \emph{and} on a choice of background. This dependence was put forward by Bershadsky, Cecotti, Ooguri and Vafa (BCOV) in Refs.~\cite{Bershadsky:1993ta,Bershadsky:1993cx} and interpreted as a change of polarization of a wave function in Ref.~\cite{Witten:1993ed}. The wave function interpretation gives a background independent meaning to the topological string partition function as an abstract state in a Hilbert space which is obtained from the geometric quantization of a bundle on the moduli space. This bundle carries the analogous information as the electric and magnetic variables $a$ and $a_D$ in the Seiberg-Witten setup, the change between these variables can thus be thought of as a change between conjugate symplectic Darboux coordinates.

The wave function property is also another manifestation of the fact that the partition function reflects the physical dualities of the target space.  Whenever the duality group has an $SL(2,\mathbbm{Z})$ or a subgroup thereof sitting in it, the perturbative topological string amplitudes can be expressed as polynomials of the corresponding quasi modular forms \cite{Kaneko:1995,Dijkgraaf:1995} as shown by BCOV \cite{Bershadsky:1993cx} and many subsequent works, e.g.~\cite{Minahan:1998vr,Hosono:1999qc,Hosono:2002xj,Klemm:2004km,Huang:2006si,Aganagic:2006wq,Grimm:2007tm}. Expressing the higher genus topological string amplitudes as polynomials in the generators of the ring of quasi modular forms is in particular useful to examine the global properties of these functions \cite{Huang:2006si,Aganagic:2006wq}. In particular, the singular behavior of the higher genus free energies associated with Seiberg-Witten theory at the locus where the monopole becomes massless was used in Ref.~\cite{Huang:2006si} to fix the holomorphic ambiguity of the holomorphic anomaly recursion.

In general the differential ring of quasi modular forms for an
arbitrary target space duality group is not known. It was nevertheless
possible to prove that the higher genus topological string amplitudes
can be expressed in terms of polynomials in a finite number of
generators using the special geometry of the deformation space
\cite{Yamaguchi:2004bt,Alim:2007qj}, the structure and the freedom in
the choice of generators was further discussed in
Refs.~\cite{Alim:2008kp,Hosono:2008ve}. In terms of the
polynomial generators one can obtain global expressions for the
topological string amplitudes. In particular, the expected physical
behavior of these amplitudes at special loci in the moduli space can be examined and used for the higher genus computation on compact CYs \cite{Huang:2006hq,Haghighat:2009nr,Alim:2012ss}.

In this work we start from the differential polynomial rings as defined in Ref.~\cite{Alim:2007qj} and show that special choices of the generators as well as of the coordinate on the moduli space lead to a special form of the polynomial ring which allows us to define a grading with nice properties.
For certain families of non-compact CY threefolds, we identify the
moduli spaces of complex structures with some modular curves and
explore their arithmetic properties. The generators and the grading
coincide with the generators of the ring of quasi modular forms and
the modular weight for CY families with duality groups for which these
forms are known explicitly. They provide a generalization thereof for
the other cases. Having expressed the topological string amplitudes in
terms of quasi modular forms of the duality group it is furthermore
possible to act on the amplitudes with an involution, the Fricke
involution, on the moduli space which exchanges the expansion of the
quasi modular forms at two different cusps. We
interpret this involution as a duality operation which exchanges the
large complex structure and the conifold loci, providing the analogs of the action of electric-magnetic or $\mathcal{N}=2$ gauge theory $S-$duality, in the sense put forward by Ref.~\cite{Seiberg:1994rs} and examined in much more detail recently following Refs.~\cite{Argyres:2007cn,Gaiotto:2009we}.

The plan of this paper is as follows. In Section~\ref{sec:polrings} we start by reviewing the setup and the definition of the differential polynomial rings which are defined using the special geometry of the moduli space of a CY. The BCOV anomaly equations lead to a polynomial recursion for the topological string amplitudes. We work out the polynomial structure in special coordinates for the functions obtained from the amplitudes which are sections of different powers of a line bundle on the moduli space. We proceed by defining a modified set of generators as well as a new coordinate $\tau$ and define the special polynomial ring obtained in this way and discuss its properties.

In Section~\ref{sec:duality} we review elements of the theory of quasi modular forms. In particular we review the construction of the appropriate rings of quasi modular forms for the subgroups $\Gamma_0(N), N=1^*,2,3,4$ of the full modular group $PSL(2,\mathbbm{Z})$.\footnote{The notation $1^*$ is introduced in Sec.~\ref{subsec:modular}} We highlight the Fricke involution which acts on the generators of the rings of quasi modular forms and exchanges their expansions at two different cusps of the modular curves. Relating these moduli curves to the moduli spaces of non-compact CY manifolds, we interpret the action of the involution as an action of a duality which exchanges the large complex structure and conifold loci.

In Section~\ref{sec:applications} we construct the special polynomial rings defined in Section~\ref{sec:polrings} for a number of examples. In the case of non-compact CY geometries, which we study on the B-side, the special polynomial rings coincide with the rings of quasi modular forms when the duality group is $\Gamma_0(N)$ for $N=1^*,2,3,4$. We study local $\mathbbm{P}^2$, and local geometries with $E_n$ singularities for $n=5,6,7,8$. We verify that the duality action exchanges the two different expansion loci in the moduli space. We furthermore apply the general construction of the special polynomial rings to the case of the quintic, this differential polynomial ring provides a generalization of the rings of quasi modular forms for this case. Finally, we give our conclusions and provide some technicalities in the appendices.


\section{Special geometry polynomial rings}\label{sec:polrings}
\subsection{Special geometry ring}
Special geometry\footnote{For a review see \cite{Ceresole:1993qq} and references therein.} is the target space geometric realization of the chiral ring \cite{Lerche:1989uy} underlying mirror symmetry. It describes the geometry of the moduli space $\mathcal{M}$ using the variation of a decomposition  of a bundle $\mathcal{H}$ over  $\mathcal{M}$. While this structure is common to both the A- and B-sides of mirror symmetry we will adopt the language of the B-model in the following. Fixing a complex structure on $\mathcal{M}$, at any given point in the base manifold the fibre of the bundle $\mathcal{H}$ can be decomposed into the following form \cite{Bershadsky:1993cx} (For the B-model this gives the familiar Hodge decomposition):
\begin{equation}\label{bundle}
\mathcal{H}=\mathcal{L} \oplus \mathcal{L}\otimes T\mathcal{M}\oplus \overline{\mathcal{L}}\otimes \overline{T\mathcal{M}} \oplus \overline{\mathcal{L}}\,,
 \end{equation}
 where $\mathcal{L}$ is the Hodge line bundle, $T\mathcal{M}$ is the holomorphic tangent
 bundle and $\overline{\mathcal{L}}$ and $\overline{T\mathcal{M}}$ are
 their complex conjugates. An additional ingredient is the cubic
 coupling which is a holomorphic section of $\mathcal{L}^2\otimes \textrm{Sym}^3T^*\mathcal{M}$, these are denoted by $C_{ijk}, i,j,k=1,\dots,n=\textrm{dim}\mathcal{M}$. The metric on $\mathcal{L}$ is denoted by $e^{-K}$ and provides a K\"ahler potential for a metric on $\mathcal{M}$ given by $G_{i\bar{\jmath}}=\partial_i\partial_{\bar{\jmath}}K$. The curvature of the metric is furthermore given by:
 \begin{equation}
R_{i\bar{\imath}\phantom{l}j}^{\phantom{i\bar{\imath}}l}=[\bar{\partial}_{\bar{\imath}},D_i]^l_{\phantom{l}j}=\bar{\partial}_{\bar{\imath}} \Gamma^l_{ij}= \delta_i^l
G_{j\bar{\imath}} + \delta_j^l G_{i\bar{\imath}} - C_{ijk} \overline{C}^{kl}_{\bar{\imath}},
\label{curvature}
 \end{equation}
where $D_i$ is the covariant derivative with connection parts which follow from the context
\begin{equation}
\Gamma^k_{ij} =G^{k\overline{k}} \partial_i G_{\overline{k} j},  \quad\textrm{and} \quad K_i=\partial_i K,
\end{equation}
for the tangent bundle and the line bundle respectively and
\begin{equation}
\overline{C}_{\bar{\imath}}^{jk}:= e^{2K} G^{k\bar{k}} G^{l\bar{l}}\overline{C}_{\bar{\imath}\bar{k}\bar{l}}.
\end{equation}
Choosing a section $\Omega$ of $\mathcal{L}$ one can obtain sections of the summands in Eq.~(\ref{bundle}) by acting on $\Omega$ with the covariant derivatives. The $C_{ijk}$ furthermore define a ring structure on sections of $\mathcal{H}$. This can be phrased for example as
\begin{equation}
D_i D_j \Omega=i C_{ijk} e^K G^{k\bar{k}} D_{\bar{k}}\overline{\Omega}\,.
\end{equation}

\subsection{Holomorphic anomaly and polynomial ring}

\subsubsection*{\it Holomorphic anomaly equations}
The topological string amplitude or free energy $\F{g}$ at genus $g$ as defined in Ref.~\cite{Bershadsky:1993cx} is a section of the line bundle
$\mathcal{L}^{2-2g}$ over $\mathcal M$. The correlation function at genus $g$ with $n$
insertions $\mathcal{F}^{(g)}_{i_1\cdots i_n}$ is only non-vanishing for
$(2g-2+n)>0$. They are related by taking covariant derivatives as this
represents insertions of chiral operators in the bulk, e.g.~$D_i \mathcal{F}^{(g)}_{i_1\cdots i_n}=\mathcal{F}^{(g)}_{ii_1\cdots i_n}$.

In \cite{Bershadsky:1993cx} it is shown that the genus $g$ amplitudes
are recursively related to lower genus amplitudes by the holomorphic anomaly equations:
\begin{equation}
\bar{\partial}_{\bar{\imath}} \F{g} = \h \overline{C}_{\bar{\imath}}^{jk} \left(
\sum_{r=1}^{g-1}
D_j\mathcal{F}^{(r)} D_k\mathcal{F}^{(g-r)} +
D_jD_k\mathcal{F}^{(g-1)} \right) \label{anom1}.
\end{equation}

At genus $1$, an additional equation was given in Ref.~\cite{Bershadsky:1993ta}
\begin{equation}
\bar{\partial}_{\bar{\imath}} \mathcal{F}^{(1)}_j = \frac{1}{2} C_{jkl}
\overline{C}^{kl}_{\bar{\imath}}+ (1-\frac{\chi}{24})
G_{j \bar{\imath}}\,, \label{anom2}
\end{equation}
 where $\chi$ is the Euler character of the CY threefold.
All higher genus $\mathcal{F}^{(g)}$ are determined recursively from these up to a holomorphic ambiguity.

A solution of the recursion equations is given in terms of Feynman rules \cite{Bershadsky:1993cx}.
The propagators $S$, $S^i$, $S^{ij}$ for these Feynman rules are related
to the three-point or Yukawa couplings $C_{ijk}$ as
\begin{equation}
\partial_{\bar{\imath}} S^{ij}= \overline{C}_{\bar{\imath}}^{ij}, \qquad
\partial_{\bar{\imath}} S^j = G_{i\bar{\imath}} S^{ij}, \qquad
\partial_{\bar{\imath}} S = G_{i \bar{\imath}} S^i.
\label{prop}
\end{equation}
By definition, the propagators $S$, $S^i$ and $S^{ij}$ are sections of
the bundles
$\mathcal{L}^{-2}\otimes \text{Sym}^m T\mathcal{M}$ with $m=0,1,2$, respectively.
The vertices of the Feynman rules are
given by the correlation functions $\mathcal{F}^{(g)}_{i_1\cdots
i_n}$.

\subsubsection*{\it Polynomial structure}
In Ref.~\cite{Yamaguchi:2004bt} it was proven that the higher genus topological string amplitudes for the quintic and related CY families with one-dimensional moduli spaces can be expressed as polynomials in a finite number of generators obtained from the closure of the ring of multi-derivatives acting on the connections. In Ref.~\cite{Alim:2007qj}, the generalization of this construction was given for an arbitrary CY. It was proven that the correlation functions
$\F{g}_{i_1\cdots i_n}$ are polynomials of degree $3g-3+n$ in the
generators $K_i,S^{ij},S^{i},S$ where a grading $1,1,2,3$ was assigned to these generators respectively. It was furthermore shown that by making a change of generators \cite{Alim:2007qj}
\begin{eqnarray}\label{shift}
\tilde{S}^{ij} &=& S^{ij}, \nonumber \\
\tilde{S}^i &=& S^i - S^{ij} K_j, \nonumber \\
\tilde{S} &=& S- S^i K_i + \frac{1}{2} S^{ij} K_i K_j, \nonumber\\
\tilde{K}_i&=& K_i\, ,
\end{eqnarray}
the $\mathcal{F}^{(g)}$ do not depend on $\tilde{K}_i$, i.e. $\partial \mathcal{F}^{(g)}/\partial \tilde{K}_i=0$.

The proof is inductive and starts by expressing the first non-vanishing correlation functions in terms of these generators. At genus zero these are
the holomorphic three-point couplings $\mathcal{F}^{(0)}_{ijk} = C_{ijk}$.
The holomorphic anomaly equation Eq.~(\ref{anom1}) can be integrated
using Eq.~(\ref{prop}) to
\begin{equation}
\mathcal{F}^{(1)}_i = \h C_{ijk} S^{jk} +(1-\frac{\chi}{24}) K_i
+ f_i^{(1)}, \label{sol2}
\end{equation}
with ambiguity $f_i^{(1)}$. As can be seen from
this expression, the non-holomorphicity of the correlation functions
only comes from the generators. Furthermore the special geometry relation (\ref{curvature}) can be integrated:
\begin{equation}
\Gamma^l_{ij} = \delta_i^l K_j + \delta^l_j K_i - C_{ijk} S^{kl} + s^l_{ij}\,,
\label{specgeom}
\end{equation}
where $s^l_{ij}$ denote holomorphic functions that are not fixed by the
special geometry relation\footnote{See Ref.~\cite{Alim:2008kp} for a discussion of how many of these are independent.}, this can be used to derive the following equations which show the closure of the generators carrying the non-holomorphicity under taking derivatives \cite{Alim:2007qj}.\footnote{These equations are for the tilded generators of Eq.(\ref{shift}) and are obtained straightforwardly from the equations in Ref.~\cite{Alim:2007qj}.}
\begin{eqnarray} \label{rel}
D_i \tilde{S}^{jk} &=&\delta_i^j(\tilde{S}^k+\tilde{S}^{km}K_m) +\delta_i^k(\tilde{S}^j+\tilde{S}^{jm}K_m) -C_{imn} S^{mj} S^{nk}  + h_i^{jk} \, , \nonumber\\
D_i \tilde{S}^j &=& 2\delta^j_i\tilde{S} + \delta^{j}_i\tilde{S}^mK_m-\tilde{S}^jK_i-k_{ik}\tilde{S}^{jk}+h_i^j\,,\nonumber\\
D_i \tilde{S} &=& \frac{1}{2} C_{imn} \tilde{S}^m \tilde{S}^n-2K_i\tilde{S} -h_{ij} \tilde{S}^j +h_i \, ,\nonumber\\
D_i K_j &=& -K_i K_j  -C_{ijk} \tilde{S}^k + k_{ij} \, ,
\end{eqnarray}
an additional equation for the holomorphic $C_{ijk}$ can be derived in a similar fashion:
\begin{eqnarray}
D_{i}C_{jkl}&=&-K_i C_{jkl}-K_j C_{ikl}-K_k C_{jil}-K_l C_{jki}+C_{ijm}\tilde{S}^{mn}C_{nkl}\nonumber\\&&+C_{ikm}\tilde{S}^{mn}C_{njl}+C_{ilm}\tilde{S}^{mn}C_{njk}+h_{ijkl}\,,
\end{eqnarray}
where $h_i^{jk}, h^j_i$, $h_i,k_{ij}$ and $h_{ijkl}$ denote
holomorphic sections of
$\mathcal{L}^{-2},\mathcal{L}^{-2},\mathcal{L}^{-2},\mathcal{L}^{0},\mathcal{L}^{2}$
respectively. All these sections together with the functions
$s_{ij}^l$ in Eq.~(\ref{specgeom}) are not independent. It was shown
in Ref. \cite{Alim:2008kp} (see also Ref.~\cite{Hosono:2008ve}) that the
freedom of choosing the holomorphic sections in this ring reduces to holomorphic sections $\cE^{ij},\cE^j,\cE$ which can be added to the polynomial generators
\begin{eqnarray} \label{freedom}
\widehat{S}^{ij} &=& \tilde{S}^{ij} + \cE^{ij} \, ,\nonumber\\
\widehat{S}^{j} &=& \tilde{S}^{j} + \cE^{j} \, ,\nonumber\\
\widehat{S} &=& \tilde{S} + \cE \, .
\end{eqnarray}

All the holomorphic quantities change according to the following equations
\begin{eqnarray}\label{holfreedom}
\hat{s}_{ij}^k &=& s_{ij}^k + C_{ijl} \mE^{lk} \, ,\nonumber\\
\hat{k}_{ij} &=& k_{ij} + C_{ijl} \mE^{l} \, ,\nonumber\\
\hat{h}_{i}^{jk} &=& h_{i}^{jk} + \theta_i \mE^{jk} + \hat{s}_{im}^j \mE^{mk} + \hat{s}_{im}^k \mE^{mj} - C_{imn} \mE^{nk} \mE^{mj} - \delta_i^k \mE^j -\delta_i^j \mE^k \, ,\nonumber\\
\hat{h}_i^j &=& h_i^j + \theta_i \mE^j - 2 \delta_i^j \mE+\hat{s}_{im}^j \mE^m -C_{imn} \mE^{nj} \mE^m  + \hat{k}_{ik} \mE^{kj} \, ,\nonumber\\
\hat{h}_i&=& h_i + \theta_i \mE -\frac{1}{2} C_{imn} \mE^m \mE^n + \hat{k}_{ij} \mE^j \, .
\end{eqnarray}

The topological string amplitudes now satisfy the holomorphic anomaly equations where the $\bar{\partial}_{\bar{\imath}}$ derivative is replaced by derivatives with respect to the polynomial generators \cite{Alim:2007qj}.
\begin{eqnarray}\label{polrec1}
\bar{\partial}_{\bar{\imath}} \mathcal{F}^{(g)} &=&  \overline{C}_{\bar{\imath}}^{jk} \left(  \frac{\partial \mathcal{F}^{(g)}}{\partial S^{jk}} -\frac{1}{2}  \frac{\partial \mathcal{F}^{(g)}}{\partial \tilde{S}^{k}}K_j -\frac{1}{2}  \frac{\partial \mathcal{F}^{(g)}}{\partial \tilde{S}^{j}}K_k + \frac{1}{2}  \frac{\partial \mathcal{F}^{(g)}}{\partial \tilde{S}}K_jK_k\right)+G_{\bar{\imath}j} \frac{\partial \mathcal{F}^{(g)}}{\partial K_{j}} \\
&=& \frac{1}{2} \overline{C}_{\bar{i}}^{jk} \left(
\sum_{r=1}^{g-1}
D_j\mathcal{F}^{(r)} D_k\mathcal{F}^{(g-r)} +
D_jD_k\mathcal{F}^{(g-1)} \right)
\end{eqnarray}
assuming the linear independence of $\overline{C}_{\bar{\imath}}^{jk}$ and $G_{\bar{\imath}j}$, this gives two sets of equations by setting the coefficients of these functions to zero. The second set of equations in particular dictates that:
\begin{equation}
\frac{\partial \mathcal{F}^{(g)}}{\partial K_{j}} =0\,.
\end{equation}


\subsection{Boundary conditions}

The mirror pair of CY threefolds $(Z, Z^*)$ we denote by $z_i$, $i =
1, \dots, n=h^{1,1}(Z)=h^{2,1}(Z^*),$ the local coordinates on the moduli space $\cM$ of
complex structures of $Z^*$ near the large complex structure
limit. The loci in $\cM$ at which the complex structure becomes
singular are described by the components $\Delta_a(z)$ of the discriminant, where $a$ runs over the number of discriminant components.

\subsubsection*{\it Genus 1}
The holomorphic anomaly equation at genus $1$  (\ref{anom2}) can be integrated to give:
\begin{equation}
\label{genus1}
\mathcal{F}^{(1)}= \frac{1}{2} \left( 3+n -\frac{\chi}{12}\right) K +\frac{1}{2} \log \det G^{-1} +\sum_{i=1}^n s_i \log z_i + \sum_a r_a \log  \Delta_a \,.
\end{equation}
The coefficients $s_i$ and $r_a$ are fixed by the leading singular behavior of $\mathcal{F}^{(1)}$ which is given by \cite{Bershadsky:1993cx}
\begin{equation}\label{bdyconditionatlcsl}
\mathcal{F}^{(1)} \sim -\frac{1}{24} \sum_i \log z_i \int_Z c_2 J_i \, ,
\end{equation}
for a discriminant $\Delta$ corresponding to a conifold singularity the leading behavior is given by
\begin{equation}
 \mathcal{F}^{(1)} \sim -\frac{1}{12} \log \Delta \,.
 \end{equation}
All physical particles which become massless somewhere in the moduli space contribute to the genus $1$ amplitude \cite{Vafa:1995ta}, this is extremely useful in anticipating the singularities at higher genus. For instance, we will study two different types of examples regarding the singularity at the orbifold expansion point.

\subsubsection*{\it Higher genus boundary conditions}
The holomorphic ambiguity needed to reconstruct the full topological string amplitudes can be fixed by imposing various boundary conditions for $\mathcal{F}^{(g)}$ \footnote{Technically, these conditions are satisfied by the holomorphic limits of $\mathcal{F}^{(g)}$, which are defined in \cite{Bershadsky:1993cx, Bershadsky:1993ta} and recalled below in (\ref{hollimit}).}at the boundary of the moduli space.

\subsubsection*{{\it The large complex structure limit}}
The leading behavior of $\mathcal{F}^{(g)}$ at this point (which is mirror to the large volume limit) was computed in \cite{Bershadsky:1993ta,Bershadsky:1993cx,Marino:1998pg,Gopakumar:1998ii,Faber:1998,Gopakumar:1998jq}. In particular the
contribution from constant maps is
\begin{equation} \label{constmaps}
 \mathcal{F}^{(g)}|_{q_a=0}= (-1)^g \frac{\chi}{2} \frac{|B_{2g} B_{2g-2}|}{2g\,(2g-2)\,(2g-2)!} \; , \quad g>1,
\end{equation}
where $q_a$ denote the exponentiated mirror map at this point.

\subsubsection*{{\it Conifold-like loci}}

The leading singular behavior of the partition function $\mathcal{F}^{(g)}$ at a conifold locus has been determined in \cite{Bershadsky:1993ta,Bershadsky:1993cx,Ghoshal:1995wm,Antoniadis:1995zn,Gopakumar:1998ii,Gopakumar:1998jq}
\begin{equation} \label{Gap}
 \mathcal{F}^{(g)}(t_c)= b \frac{B_{2g}}{2g (2g-2) t_c^{2g-2}} + O(t^0_c),
\qquad g>1\,,
\end{equation}
Here $t_c\sim \Delta^{\frac{1}{m}}$ is the flat coordinate
at the discriminant locus $\Delta=0$. For a conifold singularity $b=1$ and $m=1$. In particular the leading singularity in \eqref{Gap} as well as the absence of subleading singular terms follows from the Schwinger loop computation of \cite{Gopakumar:1998ii,Gopakumar:1998jq}, which computes the effect of the extra massless hypermultiplet
in the space-time theory \cite{Vafa:1995ta}. The singular structure and the ``gap''
of subleading singular terms have been also observed in the dual matrix model
\cite{Aganagic:2002wv} and were first used in \cite{Huang:2006si,Huang:2006hq}
to fix the holomorphic ambiguity at higher genus. The space-time derivation of \cite{Gopakumar:1998ii,Gopakumar:1998jq} is
not restricted to the conifold case and applies also to the case $m>1$
singularities which give rise to a different spectrum of
extra massless vector and hypermultiplets in space-time.
The coefficient of the Schwinger loop integral is a weighted trace over the spin of the particles~\cite{Vafa:1995ta, Antoniadis:1995zn} leading to the prediction $b=n_H-n_V$ for the coefficient of the leading singular term.

\subsubsection*{{\it The holomorphic ambiguity}}
\label{sec:it-holom-ambig}

\def\tz{\tilde{z}}
The singular behavior of $\mathcal{F}^{(g)}$ is taken into account by the local ansatz
\begin{equation}\label{ansatzha}
\mathrm{hol. ambiguity}\sim \frac{p(z_i)}{\Delta^{(2g-2)}},
\end{equation}
for the holomorphic ambiguity near $\Delta=0$,
where $p(z_i)$ is generically a polynomial in $z$ obtained by combining the local information at the various boundary points of the moduli space.


\subsection{Special coordinates}

A special set of coordinates on the moduli space of complex structures
of $Z^*$ will be discussed which permit an identification with the physical deformations of the underlying theory (see Ref.~\cite{Ceresole:1993qq} and references therein). The discussion in the following is on the B-model side, the special coordinates defined here provide the mirror maps which give the local coordinates on the A-side. Choosing a symplectic basis of 3-cycles $A^I,B_J \in H_3(Z^*)$
\begin{equation}
A^I\cap B_J=\delta^I_J=-B_J\cap A^I\,,\quad  A^I\cap A^J=B_I\cap B_J=0\, \, ,\nonumber\\
\end{equation}
and a dual basis $\alpha_I,\beta^J$ of $H^3(Z^*)$ such that
\begin{equation}
\int_{A^I} \alpha_J =\delta^I_J\, , \quad\int_{B_J} \beta^I =\delta^I_J\, ,\quad  I,J=0,\dots h^{2,1}(Z^*)\, ,
\end{equation}
the $(3,0)$ form $\Omega(x)$ can be expanded in the basis $\alpha_I,\beta^J$:
\begin{equation}
\Omega(x)= X^I(x) \alpha_I + \mathcal{F}_J(x) \beta^J\, .
\end{equation}
The periods $X^I(x),\mathcal{F}_J(x)$ satisfy the Picard--Fuchs equation of the  B-model CY family and can be identified with projective coordinates on $\mathcal{M}$ and $\mathcal{F}_J$ with derivatives of a homogeneous function $\mathcal{F}(X^I)$ of weight 2 such that $\mathcal{F}_J=\frac{\partial \mathcal{F}(X^I)}{\partial X^J}$. In a patch where $X^0(x)\ne 0$ a set of special coordinates can be defined
\begin{equation} \label{special}
t^a=\frac{X^a}{X^0}\, ,\quad a=1,\dots ,h^{2,1}(Z^*).
\end{equation}
The normalized holomorphic $(3,0)$ form $v_0= (X^0)^{-1} \Omega(t)$ has the expansion:
\begin{equation}
 v_0= \alpha_0 + t^a \alpha_a +\beta^b F_b(t) + (2F_0(t)-t^c F_c(t)) \beta^0\,,
\end{equation}
where $$F_0(t)= (X^0)^{-2} \mathcal{F} \quad \textrm{and} \quad F_a(t):=\partial_a F_0(t)=\frac{\partial F_0(t)}{\partial t^a}.$$
$F_0(t)$ is the prepotential. One can define
\begin{eqnarray}
v_a &=& \alpha_a+\beta^b F_{ab}(t)+(F_a(t)-t^bF_{ab}(t))\beta^0\, ,\\
v_D^a &=& \beta^a -t^a \beta^0\, ,\\
v^0 &=& -\beta^0\, .
\end{eqnarray}
The Yukawa coupling in special coordinates is given by
\begin{equation}
C_{abc}:= \partial_a \partial_b \partial_c F_0(t)=-\int_{Z^*} v_0 \wedge \partial_a\partial_b\partial_c v_0\, .
\end{equation}
Defining further the vector with $2h^{2,1}+2$ components
\begin{equation}
v=(v_0\,, \quad v_a\,,\quad v_D^a\,,\quad v^0)^t\,,
\end{equation}
it satisfies by construction
\begin{equation}\label{flatconnection}
\partial_a \left( \begin{array}{c} v_0\\v_b\\v_D^b\\v^0 \end{array}\right)=\underbrace{\left( \begin{array}{cccc}
0&\delta^c_a&0&0\\
0&0&C_{abc}&0\\
0&0&0&\delta^b_a\\
0&0&0&0
 \end{array}
\right)}_{:=C_a} \, \left( \begin{array}{c} v_0\\v_c\\v_D^c\\v^0 \end{array}\right)\,,
\end{equation}
which defines the $(2h^{2,1}+2) \times (2h^{2,1}+2)$ connection matrices $C_a$, in terms of which the equation can be written in the form:
\begin{equation}\label{GaussManin}
\left(\partial_a-C_a\right) \, v=0\, ,
\end{equation}
this is a distinguishing property of the special coordinates.

\subsection{Special polynomial ring}

The form of the closure of the polynomial ring under derivatives and the polynomial grading of $\mathcal{F}^{(g)}$ suggests that this ring might lead to a generalization of the ring of quasi modular forms of Ref.~\cite{Kaneko:1995}. In the following we will modify the generators as well as the coordinates on the moduli space in order to provide the generalization of this ring. We will show later in a number of examples that these transformations indeed lead to the rings of quasi modular forms in cases where these are known. For the sake of simplicity we will first start with one moduli cases, a generalization to more moduli will be discussed elsewhere.

We will consider one-dimensional deformation spaces with three regular
singular points located at $(0,1,\infty)$ of an algebraic complex
structure modulus $\alpha$. This modulus is the more familiar
algebraic complex structure modulus $z$ at large complex structure, up
to a constant $d$ dictated by the Picard-Fuchs (PF) system. The form of the
three-point function in these cases is\footnote{In general this is $z^{3}C_{zzz}=\kappa/ (1-\alpha)^{{a_{3}\over 2a_{4}}}$ if the Picard-Fuchs operator is $a_{4} (1-\alpha)\,\theta^{4}-a_{3}\alpha \,\theta^{3}+\cdots$, see Ref.~\cite{Batyrev:1993wa}.}:

\begin{equation}\label{yukawa}
z^3\,C_{zzz}= \frac{\kappa}{1-\alpha}\,,\quad \alpha=d\,\cdot z\,,
\end{equation}
where $\kappa$ gives the classical triple intersection of the A-model geometry and $(1-\alpha)$ is the discriminant and $d$ is determined by the PF system.

\subsubsection*{\it Polynomial ring of sections in arbitrary coordinates}
The ring of polynomial generators (\ref{rel}) for one-dimensional moduli space with local coordinate $z$ becomes:
\begin{equation}
\label{rel1d}
\begin{aligned}
  D_z S^{zz} &=2\tilde{S}^z+2 S^{zz}K_z-C_{zzz} (S^{zz})^2   + h^{zz}_z \, , \\
  D_z \tilde{S}^z &= 2\tilde{S} -k_{zz} S^{zz}+  h_{z}^z\,,\\
  D_z \tilde{S} &= \frac{1}{2} C_{zzz} (\tilde{S}^z)^2-2K_z\tilde{S} -k_{zz} \tilde{S}^z +  h_z \, ,\\
  D_z K_z &= -(K_z)^2 -C_{zzz} \tilde{S}^z + k_{zz} \, ,
\end{aligned}
\end{equation}
in addition we have
\begin{equation}
D_{z}C_{zzz}=-4 K_z C_{zzz}+3 (C_{zzz})^2\tilde{S}^{zz}+ h_{zzzz}\,.
\end{equation}
as discussed before $S^{zz},\tilde{S}^z$ and $\tilde{S}$ are sections of $\mathcal{L}^{-2}$ and $C_{zzz}$ is a section of $\mathcal{L}^2$.

These equations expressing the closure of the non-holomorphic generators as well as the holomorphic three-point function under holomorphic derivatives in particular also hold if the holomorphic limit is taken. By this we mean fixing a base point, finding the canonical coordinates $(t,\bar{t})$ at that point and then treating these as independent variables and taking the limit $\bar{t}\rightarrow \infty$ as described in Ref.~\cite{Bershadsky:1993cx}. In this limit the K\"ahler potential and the metric reduce to
\begin{equation} \label{hollimit}
e^{-K}|_{\textrm{hol}}=X^0\,, \quad G_{z\bar{z}}|_{\textrm{hol}}=C_{\bar{z}} \frac{\partial t}{\partial z}\,,
\end{equation}
with the period $X^0$ and the flat coordinate $t$ introduced in Eq.~(\ref{special}) and where $C_{\bar{z}}$ is a constant.

A trivial choice of the freedom in defining the generators discussed in Refs.~\cite{Alim:2008kp,Hosono:2008ve} is such that all the generators vanish. In this limit all equations are trivial except for the last one which reflects the Picard-Fuchs equation. It was furthermore shown in explicit examples \cite{Alim:2007qj,Alim:2008kp,Hosono:2008ve,Haghighat:2009nr,Alim:2012ss,Alim:2012gq} that there are choices of $\mathcal{E}^{zz},\mathcal{E}^z,\mathcal{E}$ such that $h^{zz}_z,h^{z}_z,h_{zz},h_z,h_{zzzz}$ are rational functions in $z$ of the form $p(z)/\Delta$ where $\Delta$ is the discriminant, and $p(z)$ is a polynomial in the algebraic complex structure coordinates.

\subsubsection*{\it Polynomial ring of functions in special coordinates}
In order to obtain a ring of functions on the moduli space instead of sections of powers of the line bundle  $\mathcal{L}$, we choose $X^0$, a section of $\mathcal{L}$, to multiply the different sections and produce functions. Furthermore we switch to the special coordinates $t$ and consider the following transformed objects:
\begin{equation}
S^{tt}=(\partial_z t)^2 (X^0)^2\, S^{zz} \,,\quad \tilde{S}^t=(\partial_z t) (X^0)^2\, \tilde{S}^z\,,\quad \tilde{S}_0=(X^0)^2\, \tilde{S}\,, \quad K_t= (\partial_z t)^{-1} \, K_z\,,
\end{equation}
as well as
\begin{equation}
C_{ttt}= (X^0)^{-2}\,(\partial_z t)^{-3} C_{zzz}\,.
\end{equation}
In the holomorphic limit (\ref{hollimit}), the expressions for the connections become:
\begin{equation}
K_z=-\partial_z \log X^0\,,\quad \Gamma_{zz}^z=\frac{\partial z}{\partial t}\frac{\partial^2 t}{\partial z^2}\,,
\end{equation}
The polynomial ring in the holomorphic limit, using the special coordinates becomes:
\begin{equation}
\label{rel1dflat}
\begin{aligned}
  \partial_t S^{tt} &= 2\tilde{S}^t+2 S^{tt}K_t-C_{ttt} (S^{tt})^2   + (X^0)^2 (\partial_z t)\, h^{zz}_z \, , \\
  \partial_t \tilde{S}^t &=  2\tilde{S}_0 - (\partial_z t)^{-2} k_{zz} S^{tt}+(X^0)^2  \,h_{z}^z\,,\\
  \partial_t \tilde{S}_0 &= \frac{1}{2} C_{ttt} (\tilde{S}^t)^2-2K_t\tilde{S}_0 -(\partial_z t)^{-2} k_{zz} \tilde{S}^t +  (\partial_z t)^{-1} (X^0)^{2} \,h_z \, ,\\
  \partial_t K_t &= -(K_t)^2   -C_{ttt} \tilde{S}^t + (\partial_z t)^{-2} k_{zz} \, ,\\
  \partial_{t}C_{ttt}&=-4 K_t C_{ttt}+3 (C_{ttt})^2\ S^{tt}+
  (X^0)^{-2} (\partial_z t)^{-4}\, h_{zzzz}\,.
\end{aligned}
\end{equation}

\subsubsection*{\it Special polynomial ring in $\tau$ coordinates}
In the following we will redefine some of the generators as well as use a different coordinate given by:
\begin{equation}
\tau=\frac{1}{\kappa} \partial_t F_t \,,
\end{equation}
which gives furthermore
\begin{equation}
\frac{\partial \tau}{\partial t}=\frac{1}{\kappa} C_{ttt}\,.
\end{equation}
We define the following functions on the moduli space:
  \begin{equation}
    \begin{aligned}
      K_0 &= \kappa\, C_{ttt}^{-1} \, (\theta t)^{-3} \,,&  G_1&=\theta t\,,& K_2&=\kappa\,C_{ttt}^{-1}K_t\,,\nonumber\\
      T_2&=S^{tt}\,,& T_4&= C_{ttt}^{-1} \tilde{S}^t\,,&
      T_6&=C_{ttt}^{-2} \tilde{S}_0\,,
    \end{aligned}
\end{equation}
we furthermore define holomorphic functions $\tilde{h}^{z}_{zz},\tilde{h}_{zz},\tilde{h}_z$ and $\tilde{h}^z_{zzz}$ out of the holomorphic sections $h^{zz}_z,h^{z}_z,h_z$ and $h_{zzzz}$ appearing in (\ref{rel}) in the following way:
\begin{equation}\label{specialgenerators}
\tilde{h}^z_{zz}= C_{*zz}\, h^{zz}_{z}\,,\quad \tilde{h}_{zz}= C_{*zz}\,h^{z}_z\,\quad \tilde{h}_z=C_{***}\, h_z\,,\quad \tilde{h}^{z}_{zzz}= (C_{*}^{-1})^{zz} \, h_{zzzz}\,
\end{equation}
where the star in $C_{*zz}$ denotes a fixed modulus, although in the one modulus case the distinction is not relevant it is made here to clarify the index structure of the new objects. We will furthermore redefine all objects carrying indices of the algebraic coordinate $z$ by multiplying (dividing) by $z$ for lower (upper) indices, i.e.
\begin{equation}
\tilde{h}^z_{zz}\rightarrow z\,\tilde{h}^z_{zz}\,, \quad \tilde{h}_{zz} \rightarrow z^2\, \tilde{h}_{zz}\,,\quad \tilde{h}_{z}\rightarrow z \tilde{h}_{z}\,, \quad \tilde{h}^z_{zzz} \rightarrow z^2\, \tilde{h}^z_{zzz}\,, \quad k_{zz}\rightarrow z^2 \,k_{zz}\,, \quad s_{zz}^z\rightarrow z\,s_{zz}^z
\end{equation}

The ring (\ref{rel1dflat}) using the new generators and the $\tau$ coordinate becomes:
\begin{equation}
\label{relspecial}
\begin{aligned}
\partial_{\tau} K_0&=-2K_0\,K_2- K_0^2\, G_1^2\,(\tilde{h}^z_{zzz}+3(s_{zz}^z+1))\,,\\
\partial_{\tau} G_1&= 2G_1\,K_2-\kappa  G_1\,T_2\,+K_0 G_1^3(s_{zz}^z+1)\,,\\
\partial_{\tau} K_2&=3K_2^2-3\kappa K_2\,T_2-\kappa^2 T_4+K_0^2\,G_1^4 k_{zz}-K_0\,G_1^2\,K_2\,\tilde{h}^z_{zzz}\,,\\
\partial_{\tau} T_2&=2K_2\,T_2-\kappa T_2^2+2\kappa T_4+\frac{1}{\kappa}\,K_0^2 G_1^4 \tilde{h}^z_{zz}\,,\\
\partial_{\tau} T_4&=4 K_2 T_4-3\kappa T_2\,T_4+ 2\kappa T_6-K_0\, G_1^2 \, T_4 \tilde{h}^z_{zzz}-\frac{1}{\kappa} K_0^2\, G_1^4 \,T_2 k_{zz}+\frac{1}{\kappa^2} K_0^3\, G_1^6 \tilde{h}_{zz}\,,\\
\partial_{\tau} T_6&= 6 K_2\, T_6-6\kappa T_2 \,T_6+\frac{\kappa}{2} T_4^2-\frac{1}{\kappa} K_0^2\, G_1^4 \,T_4\,k_{zz}+\frac{1}{\kappa^3} K_0^4\, G_1^8 \tilde{h}_z-2 \, K_0\,G_1^2\,T_6  \tilde{h}^z_{zzz}\,.
\end{aligned}
\end{equation}
Assuming that all remaining holomorphic functions can be expressed as rational functions in the algebraic modulus\footnote{This assumption is true in all known examples but not proven in general, see for example the discussion in Ref.~\cite{Hosono:2008ve}. Later we will see that this is well motivated from the theory of quasi modular forms, since all modular invariant functions can be expressed as rational functions of the so-called Hauptmodul, which is the generator of the function field of the moduli space and is intuitively the algebraic modulus, the quasi modular forms which require a non-holomorphic completion correspond to the elements of the ring (\ref{relspecial}).} we need a further generator in order to parameterize these. We choose a geometric object giving a function on the moduli space:
\begin{equation}\label{ratgen}
C_0=\theta \log z^3 C_{zzz}=\frac{\alpha}{1-\alpha}\,,
\end{equation}
the derivative of this generator is computed to be:
\begin{equation}
\partial_{\tau} C_0= K_0\, G_1^2\, C_0\,(C_0+1)\,.
\end{equation}
To obtain functions out of the topological string amplitudes we introduce:
\begin{equation}
F^{(g)}= (X^0)^{2g-2}\,\F{g}\,.
\end{equation}
The polynomial recursion (\ref{polrec1}) for these functions in the generators~(\ref{specialgenerators}) for the function becomes:
\begin{equation}\label{specialrec1}
\frac{\partial F^{(g)}}{\partial T_2}-\frac{1}{\kappa}\frac{\partial F^{(g)}}{\partial T_4} \,K_2 +\frac{1}{2\kappa^2} \frac{\partial F^{(g)}}{\partial T_6} K_2^2 =\frac{1}{2} \sum_{r=1}^{g-1} \partial_t F^{(g-r)}\,\partial_t\, F^{(r)} +\frac{1}{2} \partial_t^2 F^{(g-1)}\,,
\end{equation}
and
\begin{equation}\label{specialrec2}
\frac{\partial F^{(g)}}{\partial K_2}=0\,.
\end{equation}
The $t$ derivative in Eq.~(\ref{specialrec1}) can be replaced by:
\begin{equation}
\partial_t= K_0^{-1} G_1^{-3} \, \partial_{\tau}\,.
\end{equation}
The definitions (\ref{specialgenerators}), the form of the differential ring (\ref{relspecial}) and the polynomial recursion (\ref{specialrec1},\,\ref{specialrec2}) immediately lead to the following proposition, assuming that there exists a choice of $\tilde{h}^{z}_{zz},\tilde{h}_{zz},\tilde{h}_z$ and $\tilde{h}^z_{zzz}$ and $k_{zz}$ which can be expressed as rational functions in $\alpha$.

\subsubsection*{Proposition}
\begin{enumerate}
\item The differential polynomial ring generated by the special functions $C_0,K_0,G_1,K_2,T_2,T_4$ and $T_6$ closes under derivatives w.r.t. $\tau=\frac{1}{\kappa}F_{tt}$.
\item A grading is furthermore assigned to the generators given by their subscript. The $\tau$ derivative strictly increases the grading by 2.
\item $F^{(g)}$ is a polynomial of degree zero in the generators, obtained recursively from Eqs.~(\ref{specialrec1},\,\ref{specialrec2}) up to the addition of a rational function of the form
$$ A^{(g)}= K_0^{g-1}\, P^{(g)}(C_0),$$
where $P^{(g)}$ is a rational function in $C_0$ chosen such that $F^{(g)}$ respects the boundary conditions.
\item $F^{(g)}_n=\partial_t^n F^{(g)}$ is a polynomial of degree $-n$ in the generators.
\end{enumerate}

Most of the content of the proposition follows immediately from the definitions. The grading of $F^{(g)}$ is proven recursively starting from the initial data of the recursion given by the first non-vanishing amplitudes at genus $0$:
\begin{equation}
F^{(0)}_{ttt}=C_{ttt}= \kappa\, K_0^{-1}\, G_1^{-3}\,,
\end{equation}
and at genus $1$ (\ref{sol2}):
\begin{equation}\label{genus1mod}
F_t^{(1)}=\frac{1}{2} \kappa K_0^{-1} \,G_1^{-3} T_2 + (1-\frac{\chi}{24}) K_0^{-1}\, G_1^{-3} K_2+ G_1^{-1} f^{(1)}_z.
\end{equation}
It will be later shown in examples that this grading agrees with the modular weight of quasi modular forms for special geometries with duality groups for which these are known.


\section{Quasi modular forms and duality }\label{sec:duality}
In this section we review some concepts of modular curves, modular forms and quasi
modular forms for some of the groups $\Gamma_0(N)$, which will be defined in
the sequel. We highlight an involution, the \emph{Fricke involution}
acting on the modular curves, and thus on quasi modular forms and exchanging their expansions at two
different cusps. We interpret this involution as giving the
mathematical operation which corresponds to a physical duality which is the analog of the $\mathcal{N}=2$ gauge theory electric-magnetic or S-duality in the sense of Refs.~\cite{Seiberg:1994rs,Argyres:2007cn,Gaiotto:2009we}. The physical terminology of electric
and magnetic duality is motivated from the $\mathcal{N}=2$, $4d$ duality in
Seiberg-Witten gauge theories which refers to the fact that in the
coupling space there are different expansion points where the theory
looks completely different. At a weak coupling region, the electric
degrees of freedom are relevant whereas near a point in the moduli
space where magnetic degrees of freedom become massless a different
effective theory is needed. The Seiberg-Witten duality was used in Ref.~\cite{Huang:2006si}, where the special geometry of $SU(2)$ Seiberg-Witten theory was cast in terms of quasi modular forms which
were expanded both in the weak coupling and in the magnetic regime. For an early discussion of electric magnetic duality and congruence subgroups in Seiberg-Witten theory see~Ref.~\cite{Minahan:1996ws}.

Interpreted in terms of the geometry of a CY threefold moduli spaces, the exchange of these two special expansion points gets mapped to an exchange of the large complex structure point and
the conifold locus. The Seiberg-Witten theories with different gauge groups were engineered using type IIB \cite{Klemm:1996bj} as well as type IIA compactifications on CY geometries \cite{Katz:1996fh,Katz:1997eq}. However, even for non-compact CY geometries, only a subset of these admits an interpretation in terms of a $4d$ gauge theory. In general, the $4d$ physical theory will be characterized by its BPS states which reflect the cohomology of the CY. It is furthermore known that the conifold locus corresponds to the point in moduli space where magnetic BPS states become massless, see for example Refs.~\cite{Strominger:1995cz,Vafa:1995ta}. The breakdown of the effective theory and correspondingly of the topological string free energies at the expansion locus where magnetically charges states become massless is characterized by a leading singularity discussed in (\ref{Gap}). This equation has no sub-leading singular contributions precisely when the coordinate which is used corresponds to the mass of the state in the vicinity of the locus where it becomes massless, this is captured by the flat coordinate at the conifold expansion locus. The use of this as a boundary condition to fix the holomorphic ambiguity for the topological string free energies was pioneered in Ref.~\cite{Huang:2006hq}. In order to do so, the polynomial generators of Ref.~\cite{Yamaguchi:2004bt} were expanded in various patches of moduli space exploiting their global properties. Similar computations using the generators defined in Ref.~\cite{Alim:2007qj} were done in Refs.~\cite{Haghighat:2008gw,Alim:2008kp,Haghighat:2009nr,Alim:2012ss}.

\subsection{Basic facts about modular curves and modular forms}\label{subsec:modular}

In this section we shall give a review of some basic concepts about modular curves, modular forms and quasi modular forms which will be relevant in the following, we refer to Refs.~\cite{Diamond:2005, Zagier:2008} and the references therein for more details on the basic theory.

\subsubsection*{\it Modular groups and modular curves}
The generators and relations for the group $SL(2,\mathbb{Z})$ are given by the following:
\begin{equation}
  \label{eq:monodromies}
T =
  \begin{pmatrix}
    1 & 1\\ 0 & 1
  \end{pmatrix}\,,\quad
  S=
  \begin{pmatrix}
    0& -1 \\ 1 & 0\\
  \end{pmatrix}\,,\quad S^{2}=-I\,,\quad (ST)^{3}=-I\,.
\end{equation}
We will consider in the following the genus zero congruence subgroups called Hecke subgroups of $\Gamma(1)=PSL(2,\mathbb{Z})=SL(2,\mathbb{Z})
/\{\pm I\}$
\begin{equation}
\Gamma_{0}(N)=\left\{
\left.
\begin{pmatrix}
a & b  \\
c & d
\end{pmatrix}
\right\vert\, c\equiv 0\,~ \textrm{mod} \,~ N\right\}< \Gamma(1)
\end{equation}
with $N=2,3,4$. We shall also denote $SL(2,\mathbb{Z})$ by $\overline{\Gamma}(1)$ and
the corresponding groups of $\Gamma_{0}(N)$ in $SL(2,\mathbb{Z})$ by
$\overline{\Gamma}_{0}(N)$. A further subgroup that we will consider
is the unique normal subgroup in $\Gamma(1)$ of index 2 which is often
denoted $\Gamma_0(1)^*$, this is discussed in Sec.~\ref{initialdata}. By abuse of notation, we write $N=1^*$ when
listing it together with the groups $\Gamma_0(N)$.

\begin{figure}[h!]
  \centering
  \subfloat[$\Gamma_0(1)^*$]{\includegraphics[width=60mm]{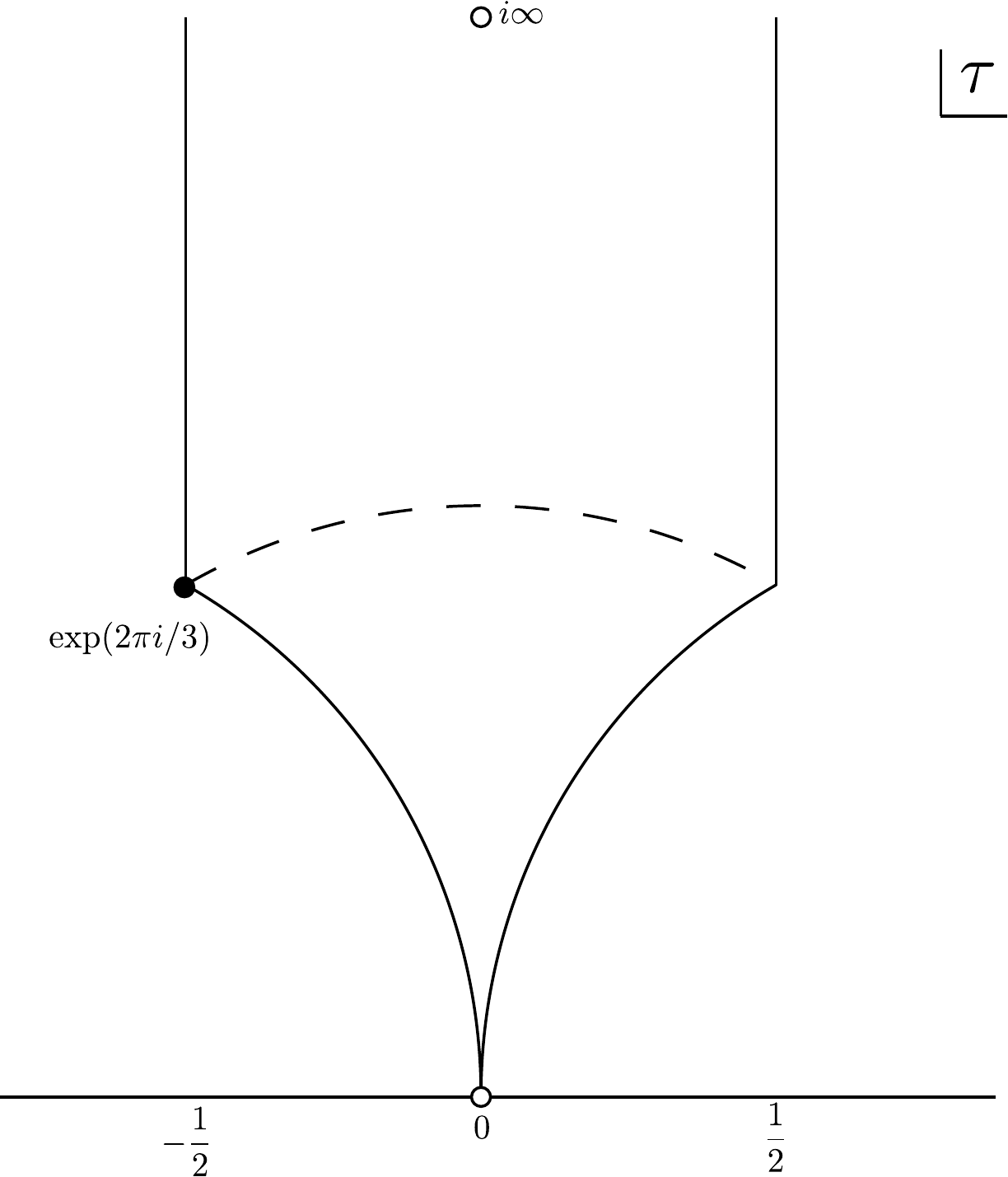}}
  \qquad
  \subfloat[$\Gamma_0(2)$]{\includegraphics[width=60mm]{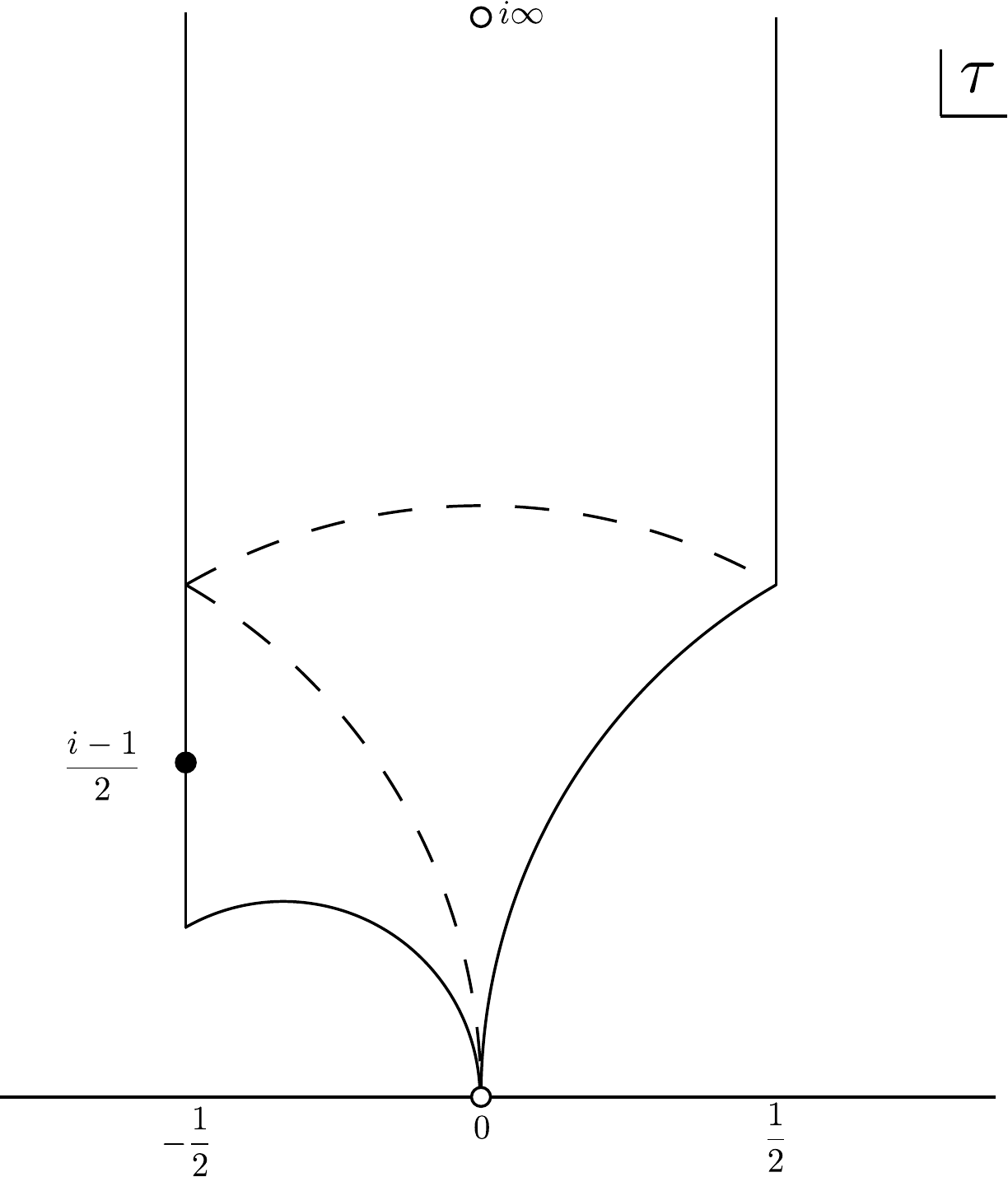}}

  \subfloat[$\Gamma_0(3)$]{\includegraphics[width=60mm]{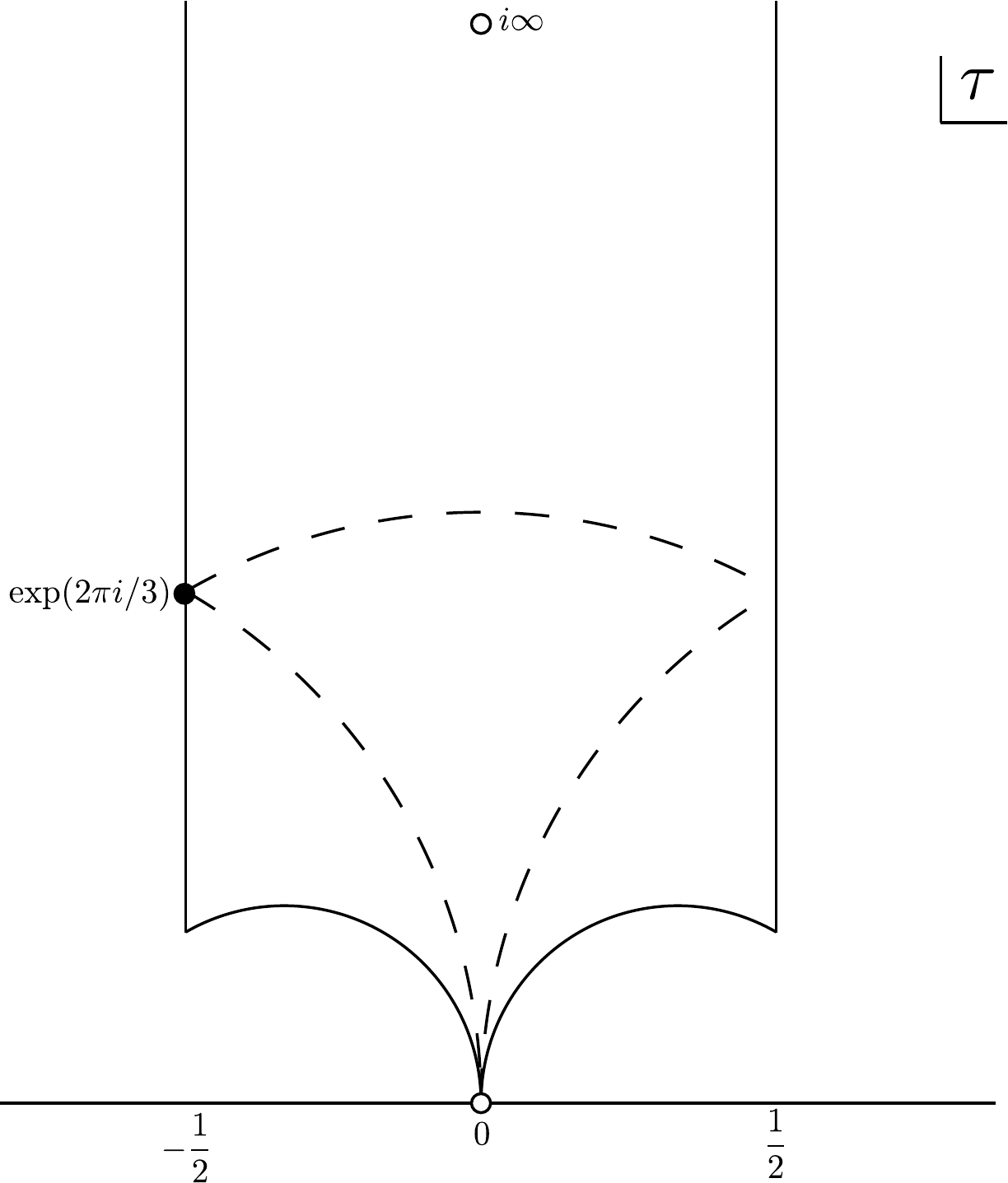}}
  \qquad
  \subfloat[$\Gamma_0(4)$]{\includegraphics[width=60mm]{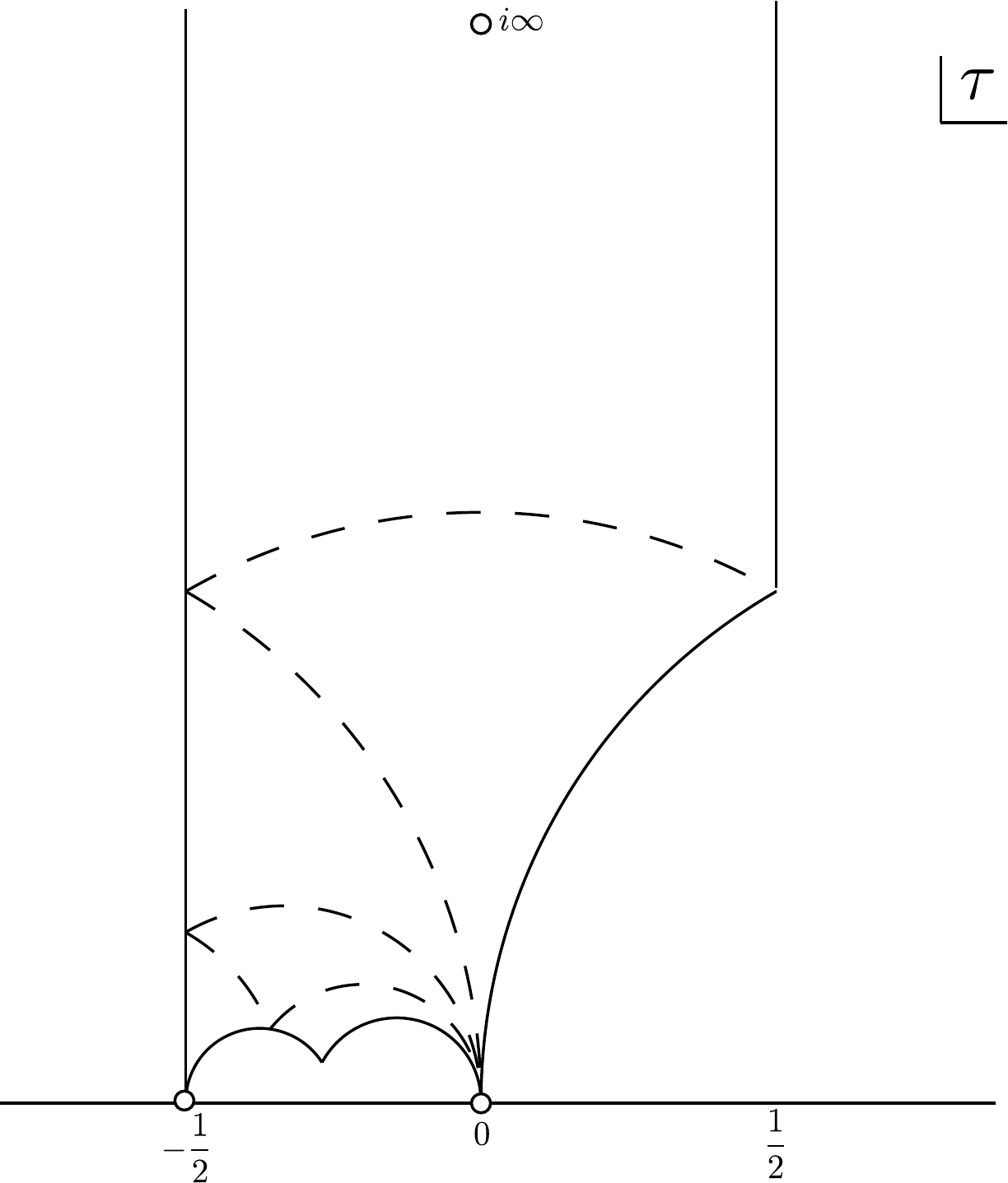}}
\caption{Fundamental domains for $\Gamma_0(1)^*$,
$\Gamma_0(N)$, $N=2,3,4$. \\The empty and full circles stand for cusps and elliptic points respectively.} \label{funddomain}
\end{figure}

The group $SL(2,\mathbb{Z})$ acts on the upper half plane $\mathcal{H}
= \{ \tau \in \mC |\, \text{Im} \tau > 0 \}$ by fractional linear
transformations:
$$\tau \mapsto \gamma\tau=\frac{a\tau+b}{c\tau+d}\quad \text{for} \quad
\gamma=\begin{pmatrix} a&b\\c&d\end{pmatrix} \in
SL(2,\mathbb{Z})\,.$$ The quotient space $Y_{0}(N)=
\Gamma_{0}(N)\backslash \mathcal{H}$ is a non-compact orbifold with
certain punctures corresponding to the cusps and orbifold points
corresponding to the elliptic points of the group $\Gamma_{0}(N)$.
By filling the punctures, one then gets a compact orbifold
$X_{0}(N)=\overline{Y_{0}(N)}=\Gamma_{0}(N)\backslash
\mathcal{H}^{*}$ where $\mathcal{H}^* = \mathcal{H} \cup \{i\infty\}
\cup \mathbb{Q}$. The orbifold $X_0(N)$ can be equipped with the
structure of a Riemann surface. The signature for the group
$\Gamma_{0}(N)$ and the two orbifolds $Y_{0}(N),X_{0}(N)$ could be
represented by $\{p,\mu;\nu_{2},\nu_{3},\nu_{\infty}\}$, where $p$
is the genus of $X_{0}(N)$, $\mu$ is the index of $\Gamma_{0}(N)$ in
$\Gamma(1)$, and $\nu_{i}$ are the numbers of
$\Gamma_{0}(N)$-equivalent elliptic fixed points or parabolic fixed
points of order $i$. The signatures for the groups $\Gamma_{0}(N)$,
$N=1^*,2,3,4$ are listed in the following table (see
e.g.~\cite{Rankin:1977ab}):
\begin{equation}\label{signature}
  \begin{array}[h]{|c|c|c|c|c|c|}
    \hline
    N & \nu_2 & \nu_3 & \nu_\infty & \mu & p\\
    \hline
    1^* & 0 & 1 & 2 & 2 & 0 \\
    2 & 1 & 0 & 2 & 3 & 0 \\
    3 & 0 & 1 & 2 & 4 & 0\\
    4 & 0 & 0 & 3 & 6 & 0\\
   \hline
  \end{array}
\end{equation}
The fundamental domains for these groups are depicted in Figure~\ref{funddomain}.

The space $X_{0}(N)$ is called a modular curve and is the moduli space
of pairs $(E,C)$, where $E$ is an elliptic curve and $C$ is a cyclic
subgroup of order $N$ of the torsion subgroup
$E_{N} \cong \mathbb{Z}_{N}^{2}$. It classifies each cyclic $N$-isogeny $\phi: E\rightarrow E/C$ up to isomorphism, see for example Refs.~\cite{Diamond:2005, Husemoller:2004} for more details.

Similarly, we can define the modular curve $X_{\Gamma}=\Gamma\backslash \mathcal{H}^{*}$ associated to a general subgroup $\Gamma$ of finite index in $\Gamma(1)$.
We refer the reader to Ref.~\cite{Diamond:2005} for more details on this.

We proceed by recalling some basic concepts in modular form theory following Ref.~\cite{Diamond:2005}.
In the following, we shall use the notation $\Gamma$ for a general subgroup of finite index in $\Gamma(1)$.
In particular, we can take $\Gamma$ to be the modular group $\Gamma_{0}(N)$ described above and discuss the modular form theory associated
to this group.

\subsubsection*{\it Modular functions}
\renewcommand{\labelenumi}{(\roman{enumi})}
A (meromorphic) modular function with respect to the a subgroup $\Gamma$ of finite index in $\Gamma(1)$ is a meromorphic function $f : X_{\Gamma}\rightarrow \mathbb{P}^{1}$.
Consider the restriction of $f$ to $Y_{\Gamma}=\Gamma\backslash \mathcal{H}$.
Since the restriction is meromorphic, we know $f$ can be lifted to
a function $f$ on $\mathcal{H}$. Then one gets a function $f: \mathcal{H}\rightarrow \mathbb{P}^{1}$
such that
\begin{enumerate}
\item $\,f(\gamma \tau)=f(\tau), \quad\forall \gamma\in \Gamma\,$.
\item $\,f$ is meromorphic on $\mathcal{H}$.
\item $\,f$ is ``meromorphic at the cusps" in the sense that the function
\begin{equation}
f|_{\gamma}: \tau\mapsto f(\gamma\tau)
\end{equation}
is meromorphic at $\tau=i\infty$ for any $\gamma\in \Gamma(1)$.
\end{enumerate}
The third condition requires more explanation. For any cusp class
$[\sigma] \in \mathcal{H}^*/\Gamma$\footnote{We use the notation
$[\tau]$ to denote the equivalence class of $\tau\in
\mathcal{H}^{*}$ under the group action of $\Gamma$ on
$\mathcal{H}^{*}$.} with respect to the modular group $\Gamma$, one
chooses a representative $\sigma\in \mathbb{Q} \cup \{i\infty\}$.
Then it is easy to see that one can find an element $\gamma\in
\Gamma(1)$ so that $\gamma: i\infty\mapsto \sigma$. Then this
condition means that the function defined by $\tau\mapsto f\circ
\gamma\,(\tau)$ is meromorphic near $\tau=i\infty$ and that the
function $f$ is declared to be ``meromorphic at the cusp $\sigma$"
if this condition is satisfied.

Therefore, equivalently, a (meromorphic) modular function
with respect to the modular group is a meromorphic function $f : \mathcal{H}\rightarrow \mathbb{P}^{1}$ satisfying the above properties on modularity, meromorphicity, and growth condition at the cusps.

\subsubsection*{\it Modular forms}
Similarly, we can define a (meromorphic) modular form of weight $k$ with respect to the group $\Gamma$ to be a (meromorphic) function $f : \mathcal{H}\rightarrow \mathbb{P}^{1}$ satisfying the following conditions:
\begin{enumerate}
\item $\,f(\gamma \tau)=j_{\gamma}(\tau)^{k}f(\tau), \quad \forall \gamma\in \Gamma\,$, where $j$ is called the automorphy factor defined by

$$j: \Gamma \times \mathcal{H}\rightarrow
\mathbb{C},\quad \left(\gamma=\left(
\begin{array}{cc}
a & b  \\
c & d
\end{array}
\right),\tau\right)\mapsto j_{\gamma}(\tau):=(c\tau+d)\,.$$
\item $\,f$ is meromorphic on $\mathcal{H}$.
\item $\,f$ is ``meromorphic at the cusps" in the sense that the function
\begin{equation}
\label{slash}
f|_{\gamma}: \tau\mapsto j_{\gamma}(\tau)^{-k} f(\gamma\tau)
\end{equation}
is meromorphic at $\tau=i\infty$ for any $\gamma\in \Gamma(1)$.
\end{enumerate}
We will need to be able to take roots of modular forms. For this
purpose one introduces a function $v: \Gamma \to \mC$, called
multiplier system of weight $k$ for $\Gamma$, such that $|v(\gamma)|=1$
and $v(\gamma_1\gamma_2) =
w(\gamma_1,\gamma_2)v(\gamma_1)v(\gamma_2)$. Here, 
$w(\gamma_1,\gamma_2)$ are numbers in $\{\pm1\}$ making
$v(\gamma)(c\tau+d)$ into an automorphy factor. 
Replacing the automorphy factor by $j_\gamma(\tau) =
v(\gamma)(c\tau+d)$ in Eq.~(\ref{slash}), one defines modular forms
with respect to a multiplier system, see for example
Ref.~\cite{Rankin:1977ab} for details.  

\subsubsection*{\it Quasi modular forms}\label{dfnofquasiform}
A (meromorphic) quasi modular form of weight $k$ with respect to the group $\Gamma$ is a (meromorphic) function $f : \mathcal{H}\rightarrow \mathbb{P}^{1}$ satisfying the following conditions:
\begin{enumerate}
\item $\,$ There exist meromorphic functions $f_{i}, i=0,1,2,3,\dots, k-1$ such that
\begin{equation}
f( \gamma\tau)=j_{\gamma}(\tau)^{k}f(\tau)+\sum_{i=0}^{k-1} c^{k-i}\,j_{\gamma}(\tau)^i  f_{i}(\tau)\,,\quad \forall  \gamma=\left(
\begin{array}{cc}
a & b  \\
c & d
\end{array}
\right)\in \Gamma\,.
\end{equation}
\item $\,f$ is meromorphic on $\mathcal{H}$.
\item $\,f$ is ``meromorphic at the cusps" in the sense that the function
\begin{equation}
f|_{\gamma}: \tau\mapsto j_{\gamma}(\tau)^{-k} f(\gamma\tau)
\end{equation}
is meromorphic at $\tau=i\infty$ for any $\gamma\in \Gamma(1)$.
\end{enumerate}

For a large class of non-compact CY threefolds, the relevant
geometry of the mirror manifolds are captured by the so-called
mirror curves \cite{Hori:2000kt}. In what follows we shall only
consider the cases where the mirror curves are elliptic curves.
These already include many interesting examples such as the mirrors
of the canonical bundle of $\mathbb{P}^{2}$ and the canonical
bundles of the del Pezzo surfaces $dP_{n},\, n=5,6,7,8$. See for
example Refs. \cite{Lerche:1996ni, Chiang:1999tz, Katz:1999xq,
Mohri:2001zz} for details.

As we shall discuss in greater detail later in Section
\ref{Encurvefamilies}, for the canonical bundle of $\mathbb{P}^{2}$
and the canonical bundles of the del Pezzo surfaces $dP_{n},\,
n=5,6,7$, the bases of the corresponding families of mirror curves
are the modular curves $X_{0}(N)$ with $N=3,4,3,2$, respectively.
The canonical bundle of $dP_{8}$ is exceptional in the sense the
base of the corresponding mirror curve family called $E_{8}$
elliptic curve family is not a modular curve of the form $X_{0}(N)$.
It is given by $\Gamma_{0}(1)^{*}\backslash\mathcal{H}^{*}$, where
$\Gamma_{0}(1)^{*}$ is the subgroup of $\Gamma(1)$ mentioned
earlier and discussed in Sec.~\ref{initialdata}. This base is a copy of $\mathbb{P}^{1}$ parametrized by a
particularly chosen coordinate $z$, and is a $2:1$ cover of the
$j$--plane $\mathbb{P}^{1}$ realized by the map $j(z)=1/ z(1-432z)$.
In the following we shall denote the base of this family of elliptic
curves by $X_{0}(1)^{*}$. See Refs.
\cite{Lian:1994zv,Klemm:1996,Maier:2009} for more discussions on
this family.

\subsection{Rings of quasi modular forms}\label{arithmetic}

In this section we show the explicit computation of the rings of quasi modular forms with respect to the groups $\Gamma_{0}(N),$ for $N=1^{*},2,3,4.$ Before introducing these we recall the familiar example of quasi modular forms for the full modular group $\Gamma(1)$, see e.g.~\cite{Zagier:2008}
and references therein for details.
\subsubsection{Quasi modular forms for the full modular group $PSL(2,\mathbbm{Z})$}
The familiar Eisenstein series $E_{4},E_{6}$ are generators of the
ring of modular forms with respect to the full modular group
$\Gamma(1)=PSL(2,\mathbbm{Z})$. The Eisenstein series $E_{2}$ is a
quasi modular form according to the definition given in
(\ref{dfnofquasiform}) since it transforms according to
\begin{equation}
E_{2}(\gamma \tau)=(c\tau+d)^{2}E_{2}(\tau)+{12\over 2\pi i}c(c\tau+d)\,, \quad\forall \gamma
=\begin{pmatrix}
a & b  \\
c & d
\end{pmatrix}
\in \Gamma(1)\,.
\end{equation}
The differential ring structure of quasi modular forms for
$\Gamma(1)$ is given by\footnote{We omit factors of $2\pi i$ in the derivatives, i.e. $\partial_\tau$  should be $\frac{1}{2\pi i} {\partial\over \partial\tau}$ throughout this work.}

\begin{equation}
\begin{aligned}
                  \partial_\tau E_{2}&={1\over 12}(E_{2}^2-E_{4})\,,\\
                  \partial_\tau E_{4}&={1\over 3}(E_{2}E_{4}-E_{6})\,,\\
\partial_\tau E_{6}&={1\over
2}(E_{2}E_{6}-E_{4}^{2})\,.
                          \end{aligned}
                          \end{equation}
The non-holomorphic function
${1\over \textrm{Im}\tau}$ transforms as
\begin{equation}
{1\over \textrm{Im}\gamma\tau}=(c\tau+d)^{2}{1\over \textrm{Im}\tau}-2ic(c\tau+d)\,.
\end{equation}
It follows that the non-holomorphic completion of $E_{2}$, which is defined by
\begin{equation}\label{dfnofEhat}
\widehat{E_{2}}(\tau,\bar{\tau})=E_{2}(\tau)-{3\over \pi \textrm{Im}\tau}\,,
\end{equation}
transforms according to
\begin{equation}\label{transformationofEhat}
\widehat{E_{2}}(\gamma \tau,\overline{\gamma \tau})=(c\tau+d)^{2}\widehat{E_{2}}(\tau,\bar{\tau})\,.
\end{equation}

\subsubsection{More general rings of quasi modular forms }
In the section we shall consider the genus $0$ modular curves
$X_{0}(N)$ with $N=1^{*},2,3,4$ and discuss the corresponding rings
of quasi modular forms. The relevant data giving the ring of quasi
modular forms as well as the modular parameter $\tau$ are captured
by the periods $\omega_0$ and $\omega_1$ of the corresponding
families of elliptic curves described later in Section
\ref{Encurvefamilies}, see also Refs.~\cite{Lian:1994zv, Klemm:1996,
Mohri:2001zz, Maier:2009,
  Maier:2011}. The periods satisfy the following Picard-Fuchs
differential equation:
\begin{equation} \label{deflc}
\mathcal{L}_c\,\omega_{i}= (\theta_{\alpha}^{2}-\alpha (\theta_{\alpha}+1/r)(\theta_{\alpha}+1-1/r))\, \omega_i=0\,,\quad i=0\,,1\,.
\end{equation}
The parameter $\alpha$ is the so-called Hauptmodul (that is, generator for the function field of the modular curve) listed
in e.g. Ref.~\cite{Maier:2009}, and $\theta_{\alpha}=\alpha{\partial \over \partial \alpha}$.
The solutions of this equation are given in terms of the hypergeometric functions:
\begin{equation}\label{hypersols}
\omega_0(\alpha) = ~_{2}F_{1}\left(1/r,1-1/r,1;\alpha\right)  \quad \textrm{and} \quad \omega_1(\alpha)=\frac{i}{\sqrt{N}} ~_{2}F_{1}\left(1/r,1-1/r,1;1-\alpha\right)\,.
\end{equation}
The numbers $r$ are give by the following:
\begin{equation*}
 \begin{array}{c|ccccc}
N&1^{*}&2&3&4&\\
r&6&4&3&2
\end{array}
\end{equation*}
They are related to the index $\mu$ of $\Gamma_0(N)$ by $r=\frac{12}{\mu}$.
The modular parameter $\tau$ is then given by:
\begin{equation}
\tau=\frac{\omega_1}{\omega_0}\,.
\end{equation}
The weight $1$ modular forms (for an appropriate multiplier system)
for these groups are listed for example in Refs.~\cite{Maier:2009,Maier:2011}. More
precisely, for the cases $N=1^{*},2,3,4$, the relevant modular forms
are given by
\begin{equation}\label{dfnofABC}
A=\omega_0\,, \quad B= (1-\alpha)^{1\over r}\,A\,,  \quad C=\alpha^{1\over r}\,A\,.
\end{equation}
Then by definition one has
\begin{equation}
A^{r}=B^{r}+C^{r}\,.
\end{equation}
By analytic continuation, it is easy to show
that as multi-valued functions on the modular curve $X_{0}(N)$ as an
orbifold, these modular forms (for a multiplier system) have divisors given by
\begin{equation}\label{asympofABC}
\textrm{Div} A={1\over r} (\alpha=\infty)\,,\quad
\textrm{Div} B={1\over r} (\alpha=1)\,,\quad
\textrm{Div} C={1\over r} (\alpha=0)\,.
\end{equation}
These will be useful later when we consider the singular behavior of
topological string amplitudes near different singular points of the
moduli space of certain Calabi-Yau threefolds.

These generators have very nice $\eta$--function expansions and arithmetic properties. For completeness, we recall the results from
\cite{Maier:2009} as follows:
\begin{equation}
 \begin{array}{c|cccc}\renewcommand{\arraystretch}{0.5}
N&A&B&C&\\
1^{*}&E_{4}(\tau)^{1\over 4}&({E_{4}(\tau)^{3\over 2}+E_{6}(\tau)\over 2})^{1\over 6}&({E_{4}(\tau)^{3\over 2}-E_{6}(\tau)\over 2})^{1\over 6}&\\
2&{(2^{6}\eta(2\tau)^{24}+\eta(\tau)^{24} )^{1\over 4} \over \eta(\tau)^2\eta(2\tau)^2}&{\eta(\tau)^{4}\over \eta(2\tau)^{2}}&2^{3 \over 2}{\eta(2\tau)^4\over \eta(\tau)^2}&\\
3&{(3^{3}\eta(3\tau)^{12}+\eta(\tau)^{12} )^{1\over 3} \over \eta(\tau)\eta(3\tau)}&{\eta(\tau)^{3}\over \eta(3\tau)}&3{\eta(3\tau)^3\over \eta(\tau)}&\\
4&{(2^{4}\eta(4\tau)^{8}+\eta(\tau)^{8} )^{1\over 2} \over \eta(2\tau)^2}=
{\eta(2\tau)^{10}\over\eta(\tau)^{4}\eta(4\tau)^{4}}&{\eta(\tau)^{4}\over \eta(2\tau)^2}&2^2{\eta(4\tau)^4\over \eta(2\tau)^2}&
\end{array}
\end{equation}
In particular, by examining the $\eta$--expansions of
the generators listed above, we get the relation $d_N\,\eta^{24}=\alpha(1-\alpha)^{9-N}A^{12}$, as pointed out in e.g. Ref.~\cite{Mohri:2001zz}, with
\begin{equation*} \label{dntable}
 \begin{array}{c|ccccc}
N&1^{*}&2&3&4&\\
d_{N}&432&64&27&16
\end{array}
\end{equation*}
There are also some nice $\theta$--function expansions for these modular forms and relations among these generators and the Eisensteins series $E_{4},E_{6}$. These are listed in the Appendix~\ref{appendixA}. Refs.~\cite{Zagier:2008, Maier:2009, Maier:2011} give more details on the arithmetic aspects.

Now we shall construct the ring of quasi modular forms and focus on the differential ring structure.
The modular forms defined in (\ref{dfnofABC}) satisfy the following equations
\begin{equation}\label{dfnofA}
\quad A^{2}=\partial_\tau\log {C^{r}\over B^{r}}={1\over
N-1}(NE_{2}(N\tau)-E_{2}(\tau))\,.
\end{equation}
These equalities are proved by direct computations using the
$\eta$--expansions of the modular forms. The first equality in the
above is equivalent to the following equation:
\begin{equation}\label{gaussschwarz}
\partial_{\tau}\alpha=\alpha(1-\alpha)A^{2}\,.
\end{equation}
Later, we shall interpret this as the
absence of instanton corrections of Yukawa couplings for elliptic curves, see for example Refs.~\cite{Lian:1994zv, Zagier:1998, Hosono:2008ve} for some related discussions on this.
To obtain the ring of quasi modular forms, we introduce the analog
of the Eisenstein series $E_{2}$ as a quasi modular form as follows:
\begin{eqnarray}\label{dfnofE}
E=\partial_\tau \log
B^{r}C^{r}\,.
\end{eqnarray}
Using the $\eta$--expansions of the
modular forms we can show
\begin{eqnarray}
E&=&{1\over N+1}(E_{2}(\tau)+NE_{2}(N\tau)),\quad N=1^{*},2,3\,,\\
E&=&{1\over 3}(4E_{2}(4\tau)+E_{2}(\tau))-{2\over 3}E_2(2\tau)\,,\quad N=4\,.
\end{eqnarray}
It follows then that  the generator $E$ transforms as a quasi
modular form under the modular groups $\Gamma_{0}(N)$ with
$N=1^{*},2,3,4$ respectively, using the above Eisenstein series
expressions. Similar to (\ref{dfnofEhat}) , one can define the
non-holomorphic completion $\widehat{E}$ of these quasi modular
forms so that they are non-holomoporhic but modular in the sense
that they transform in the way similar to
(\ref{transformationofEhat}).

From Eqs.~(\ref{dfnofABC},\,\ref{dfnofA},\,\ref{dfnofE}), one can easily deduce the differential ring structure of
quasi modular forms. The results are listed below for $N=1^{*},2,3,4$:
\begin{equation}\label{quasiring}
\begin{aligned}
\partial_\tau A&={1\over 2r}A(E+{C^{r}-B^{r}\over A^{r-2}})\,,\\
\partial_\tau B&={1\over 2r}B(E-A^{2})\,,\\
\partial_\tau C&={1\over 2r}C(E+A^{2})\,,\\
\partial_\tau E&={1\over
2r}(E^{2}-A^{4})\,.
  \end{aligned}
\end{equation}


\subsection{Duality action on topological string amplitudes}
\subsubsection{Fricke involution}
For each of the modular curves $X_{0}(N)$ with $N=1^{*},2,3,4$,  as
a covering of the $j$--plane $\Gamma(1)\backslash\mathcal{H}^{*}$,
there are three branch points. According to (\ref{signature}), they
are two distinguished cusps given by $[i\infty]=[1/N]$ and $[0]=
[1/1]$. The third branch point is a cubic elliptic point, quadratic
elliptic point, cubic elliptic point and a cusp for $N=1^*,2,3,4$,
respectively. The Fricke involution is defined by
\begin{equation}
W_{N}: \tau\mapsto -{1\over N\tau}\,.
\end{equation}
It exchanges these two cusps and fixes the third branch point, see Fig.~\ref{funddomain}.\footnote{We point out that for the Seiberg-Witten curve family
given by $y^{2}=(x^{2}-u)^{2}-\Lambda^{4}$ and with monodromy group
$\Gamma_{0}(4)$, the Fricke involution acts as $2\tau\mapsto
-{1\over 2\tau}$. In the literature, see for example
Ref.~\cite{Huang:2006si}, by redefining $\tau$ as the above $2\tau$,
the Fricke involution is realized as the $S$-transformation. }

Recall that the modular curve $X_{0}(N)$ is the moduli space of enhanced
elliptic curves $(E,C)$, where $C$ is an order $N$ subgroup of the
torsion group $E_{N}\cong \mathbb{Z}_{N}\oplus \mathbb{Z}_{N}$. Using this
interpretation, the Fricke involution acts by sending $(E,C)$ to
$(E/C,E_{N}/C)$.

It turns out from e.g. ~\cite{Maier:2009} that the Fricke
involution maps the Hauptmodul
\begin{equation}
\alpha \quad  \textrm{to} \quad \beta:=1-\alpha\,.
\end{equation}
The Fricke involution acts on the ring of quasi modular forms according to
\begin{equation}
  \begin{aligned}
    \label{fricke}
    A &\mapsto \frac{\sqrt{N}}{i} \tau\, A\,, \\
    B&\mapsto \frac{\sqrt{N}}{i}\tau \,C\,,\\
    C&\mapsto\frac{\sqrt{N}}{i}\tau B\,,\\
    E&\mapsto  N\tau^{2}E+{12\over 2\pi i}{2N\tau\over N+1},\quad
    N=1^{*},2,3\,. \\
    E&\mapsto  N\tau^{2}E+{12\over 2\pi i}{2N\tau\over 6}\,,\;\;\quad
    N=4\,.
  \end{aligned}
\end{equation}
For all cases $N=1^{*},2,3,4$,
the non-holomorphic completion $\widehat{E}(\tau,\bar{\tau})$ transforms according to:
\begin{equation}
\widehat{E}\mapsto N\tau^{2}\widehat{E}\,.
\end{equation}

 \subsubsection{CY moduli spaces as modular curves}
For a large class of non-compact CY geometries, the relevant part of
mirror geometry is captured by the mirror curve. For each of the
non-compact CY threefold geometries that we shall discuss below, the
moduli space $\mathcal{M}$ of complex structures of the non-compact
CY threefold can be identified with a modular curve $X_{0}(N)$. The
singular points on the moduli space $\mathcal{M}$ are identified
with the branch points on the modular curve thought of as an
orbifold $\Gamma_{0}(N)\backslash \mathcal{H}^{*}$. More precisely,
the large complex structure point is identified with the cusp
$[1/N]$, conifold point with $[1/1]$, and the orbifold point is the
third branch point on the modular curve, see Fig.~\ref{funddomain}.

This identification makes manifest the global meanings of the topological string amplitudes since they are now geometric objects defined on the modular curve thus globally defined. More precisely, the full non-holomorphic topological string amplitudes are
built out of the generators $A,B,C,\widehat{E}$, their holomorphic limits are quasi modular forms expressible in terms of the generators $A,B,C,E$. It also gives access to analyze the enumerative meanings of the topological string amplitudes at the other points on the moduli space, e.g., as the generating functions of orbifold Gromov-Witten invariants at the orbifold point.

The periods of the non-compact CY threefold satisfy the Picard-Fuchs
equations $\mathcal{L}_{CY} \pi = 0$. For geometries with one-dimensional moduli spaces there are three solutions to this operator, one of which is a constant (see for example Refs.~\cite{Lerche:1996ni, Chiang:1999tz,Diaconescu:1999dt}). The non-trivial periods are identified with $t$ and $F_t$. The operator $\mathcal{L}_{CY}$ furthermore factorizes, s.t.
\begin{equation}\label{factorizepf}
\mathcal{L}_{CY} \pi_i(z) = \mathcal{L}_c\, \circ \theta \,\omega_{i-1}(z_c), \quad i=1,2\,, \quad \theta=z\frac{d}{dz}\,,
\end{equation}
where $\mathcal{L}_c$ denotes the Picard-Fuchs operator associated
to a curve family and where there is a possible sign difference
between $z$ and $z_c$. By comparing their asymptotic behaviors near
the large complex structure limit, it is easy to see that one can
choose suitable normalizations for $t$ and $F_{t}$ so that
\begin{equation}
\theta \,t= \omega_0, \quad \theta F_{t}=\tau \omega_0\,.
\end{equation}

\subsubsection{Duality of topological string amplitudes}
Since the moduli space $\mathcal{M}$ of the non-compact CY threefold geometry is identified with the modular curve $X_{0}(N)$ and furthermore the action of the Fricke involution on the algebraic modulus $\alpha$ is:
\begin{equation}
W_{N}: \alpha \mapsto 1-\alpha\,,
\end{equation}
it is clear that the effect of the Fricke involution is exchanging
the large complex structure expansion point with the conifold
expansion point, and fixes the orbifold point. Note that this third
singularity is an orbifold point of the moduli space of the
non-compact CY threefolds, but it could be an orbifold point or a cusp of
the moduli space of the corresponding enhanced elliptic curves. We interpret
the action of this involution as the action of a duality which
exchanges the expansion points of the topological string amplitudes
around the large complex structure and the conifold points. By
expressing the topological string amplitudes in terms of the
generators of the ring of quasi modular forms (\ref{quasiring}) and
applying the Fricke involution (\ref{fricke}), we will check this
interpretation in Section~\ref{sec:applications} in a number of
examples.

\section{Applications}\label{sec:applications}
In this section we present applications of the previous ideas. We
consider a number of non-compact geometries, which have all been
studied before using different methods. We start with a detailed
discussion of local $\mathbbm{P}^2$, which denotes the canonical
bundle $\mathcal{O}(-3)\rightarrow \mathbbm{P}^2$, and its mirror.
Higher genus topolocial string amplitudes on this model have been
studied in a number of works using different techniques, see for
example Refs.~\cite{Klemm:1999gm,Katz:1999xq}.  The use of a
different set quasi modular forms for this example was considered in
Ref.~\cite{Aganagic:2006wq}, the polynomial generators of
Ref.~\cite{Alim:2007qj} were used in
Refs.~\cite{Haghighat:2008gw,Alim:2008kp} for higher genus
computations. Our new addition to these previous discussions
consists of the explicit identification of the rings of quasi
modular forms of $\Gamma_0(3)$ which are adapted to this specific
moduli space as well as their translation to the special geometry
ring of polynomial generators of Ref.~\cite{Alim:2007qj}.
Furthermore, this example serves as a testing ground for the duality
of topological amplitudes which turns out to exchange the large
complex structure and the conifold expansion loci.  The other
non-compact geometries which we consider are canonical bundles of
del Pezzo surfaces $dP_{n},\, n=5,6,7,8\,$ and their mirrors. These
were considered in the physical context of non-critical string
theories. For the purpose of our work, see Ref.~\cite{Lerche:1996ni}
and references therein. Higher genus computations using the
holomorphic anomaly equation and enumerative information from the
A-model for these geometries were considered in
Ref.~\cite{Katz:1999xq}. Finally we write down the polynomial
generators for the quintic CY and its mirror, although a precise
description of the quasi modular forms of the quintic is not known,
we can formally define and write down the analogs of the generators
used in the non-compact examples where the rings of quasi modular
forms are known. The rings which are thus provided should therefore
define the formal analogs of the rings of quasi modular forms for
these geometries, see also Ref.~\cite{Movasati:2011zz} for a recent
discussion of a ring of functions for the quintic.


\subsection{Local $\mathbbm{P}^2$}
\subsubsection{Initial data}
In order to obtain the effective triple intersection on this
geometry one should consider this non-compact geometry as the
decompactification limit of a compact one, see for example
Refs.~\cite{Klemm:1999gm,Hosono:2004jp,Alim:2008kp}. We consider the
compact geometry given by a degree 18 hypersurface in
$\mathbbm{P}_{1,1,1,6,9}$ and resolve the singularities. This is
described by the toric charge vectors:
\begin{equation}
\begin{array}{cc|cccccc}
t_1&-6&3&2&1&0&0&0\\
t_2&0&0&0&-3&1&1&1
\end{array}
\end{equation}
which describe an elliptic fibration over $\mathbbm{P}^2$. The
classical intersections are summarized in the classical piece of the
prepotential:
\begin{equation}
F_0(t)|_{\textrm{cl}}=\frac{1}{6}C_{abc}t^a\,t^b\,t^c=\frac{3}{2}t_1^3+\frac{3}{2}t_1^2\,t_2+\frac{1}{2}t_1\,t_2^2\,.
\end{equation}
Decoupling the K\"ahler parameter of the fiber should be done such that the classical volume of a four cycle in the new geometry is finite, from:
\begin{equation}
\partial_{t^2}F_0(t)|_{\textrm{cl}}=\frac{3}{2}t_1^2+t_1\, ,
\end{equation}
we see that this requires a modification of the K\"ahler classes first. The following change:
\begin{equation}
t_1\rightarrow t_1+\frac{1}{3}t_2\,,\quad t_2\rightarrow t_2\, ,
\end{equation}
gives the new classical prepotential:
\begin{equation}
F_0(t)|_{\textrm{cl}}=\frac{3}{2}t_1^3-\frac{1}{18}t_2^3\,,
\end{equation}
which now gives
\begin{equation}
\partial_{t_2}F_0(t)|_{\textrm{cl}}=-\frac{1}{6} t_2^2\,,
\end{equation}
the classical triple intersection in the CY manifold obtained from taking this limit becomes effectively
\begin{equation}
C_{ttt}|_{\textrm{cl}}=-\frac{1}{3}\,,
\end{equation}
the geometry is $\mathcal{O}(-3)\rightarrow \mathbbm{P}^2.$

\subsubsection*{Picard Fuchs equation}
A good local coordinate for the moduli space of the mirror CY manifold centered at the large complex structure point is given by\footnote{Here $z=-z_{c}$, see Ref.~\cite{Alim:2012gq} and references therein for background material.}:
\begin{equation}
z=\frac{a_1a_2a_3}{a_0^3}\,.
\end{equation}
The Picard Fuchs equation reads:
\begin{equation}
\mathcal{L}=\theta^3+3z (3\theta+1)(3\theta+2)\theta= \mathcal{L}_{c}\circ\theta\,, \quad \theta=z\frac{d}{dz}\,,
\end{equation}
the relation between the two operators is the one discussed in Section~\ref{sec:duality}. $\mathcal{L}_c$ is the operator (\ref{deflc}) for $\Gamma_0(3)$ with $\alpha=27 z_c$. The solutions of
\begin{equation}
\mathcal{L}_c=\theta^2-3z_c(3\theta+1)(3\theta+2)\,,
\end{equation}
are given in Appendix \ref{localp2}.

A full discussion of the solutions of $\mathcal{L}$ is found in Ref.~\cite{Diaconescu:1999dt}. At large complex structure these have the form
\begin{equation}
\Pi=(1\quad t \quad \frac{t^2}{2}-t+\dots)=(\pi_0(z)\quad \pi_1(z)\quad \pi_2(z))\,,
\end{equation}
such that their monodromy matrix under $t\rightarrow t+1$ is given by
\begin{equation}M_t=
\left(\begin{array}{ccc}
1&0&0\\
1&1&0\\
0&1&1
\end{array}
\right)\,.
\end{equation}
The classical piece of the period mirror to the four cycle volume is expected to be:
\begin{equation}
\partial_t F_0(t)= -\frac{1}{6} t^2\,,
\end{equation}
hence $\pi_2\sim \frac{t^2}{2}$ is identified with $-3\partial_t
F_0(t)$. From this we can obtain the prepotential.

\subsubsection*{\it Relation to periods of the curve}
The relation to the periods $\omega_0,\omega_1$ of the curve is obtained by
\begin{equation}
\theta \pi_1|_{z=-z_c}=\omega_0(z_c)\,, \quad \theta \pi_2|_{z=-z_c}=\omega_1(z_c)\,,
\end{equation}
which gives in particular
\begin{equation}
\tau=\frac{\theta \pi_2}{\theta \pi_1}|_{z=-z_c}=\frac{\partial \pi_d}{\partial t}=-3F_{tt}\,,
\end{equation}
this leads to
\begin{equation}
\partial_t = \frac{\partial \tau}{\partial t} \partial_{\tau}=-3C_{ttt}\partial_\tau\,.
\end{equation}

\subsubsection*{\it Moduli space as a modular curve}

Following the mirror construction of Ref.~\cite{Hori:2000kt}, the
family of mirror curves is given in the following form
\begin{equation}
x+1-z_{c}{x^{3}\over y}+y=0\,,\quad
j(z_{c})={(1-24z_{c})^{3}\over z_{c}^{3}(1-27z_{c})}\,.
\end{equation}
This is the Hessian family, see e.g. \cite{Husemoller:2004}. It is 3-isogenous to the other version of the Hessian family in homogeneous coordinates
\begin{equation}
x_{1}^{3}+x_{2}^{3}+x_{3}^{3}-z_{H}^{-1/3} x_{1}x_{2}x_{3}=0\,,\quad j(z_{H})={(1+216z_{H})^{3}\over z_{H}(1-27z_{H})^{3}}\,,\quad
\mathcal{L}_{H}=\theta_{z_{H}}^{2}-3z_{H}(3\theta_{z_{H}}+1)(3\theta_{z_{H}}+2)\,.
\end{equation}

These two families are not isomorphic, but they share the same base as the modular curve $X_{0}(3)$.
Interestingly, if we denote $\alpha_{H}=27z_{H}$, then it is related
to $\alpha=27z_{c}$ by the Fricke involution, due to the fact that
these two families of elliptic curves are 3-isogenous. See Refs. \cite{Chiang:1999tz, Husemoller:2004, Haghighat:2008gw} for some related discussions on these two families.

\subsubsection{Special polynomial ring as ring of quasi modular forms}
To fix the special polynomial ring we choose the following rational functions in $z$ in the construction of the ring (\ref{relspecial})\footnote{Multiplying (dividing) lower(upper) indices by $z$.}
\begin{equation}
s_{zz}^z=-\frac{4}{3}+\frac{1}{6\Delta}\,,\quad \tilde{h}^z_{zz}=\frac{1}{36\Delta^2}\,,\quad \tilde{h}^z_{zzz}=\frac{1}{2\Delta}\,,
\end{equation}
with $\Delta=1+27z$. In the notation of Sec.~\ref{sec:polrings}, the generators are $T_2,G_{1}$ and $C_0$, their relation to the generators of the differential ring of quasi modular forms of Sec.~\ref{sec:duality}  is
\begin{equation}
T_2=\frac{E}{2},\quad  G_1=A \quad \textrm{and} \quad C_0=\frac{C^3}{B^3}.
\end{equation}
We obtain the following ring
\begin{align}
&  \partial_{\tau} C_0= G_1^2\,C_0\,,\\
&  \partial_{\tau} G_1=\frac{1}{6}\left(2G_1\, T_2+G_1^3\left(\frac{C_0-1}{C_0+1}\right)\right)\,,\\
& \partial_{\tau} T_2=\frac{T_2^2}{3}-\frac{G_1^4}{12}\,.
\end{align}
The holomorphic limit of $T_4,T_6$ and $K_2$ vanishes with this choice. Furthermore since $X^0=1$ for non-compact geometries we get $K_0=(1+C_0)^{-1}$. The algebraic coordinate $z_c$ is expressed as
\begin{equation}
z_c=\frac{1}{27} \frac{C_0}{1+C_0}\,.
\end{equation}

\subsubsection*{\it Genus 1}
The genus 1 amplitude is found to be
\begin{equation}
F^{(1)}=-\frac{1}{12} \log\left((\theta t)^6 z(1+27z) \right)= -\frac{1}{12} \log\left(\frac{C_0\, G_1^6}{27(1+C_0)^2} \right)\,.
\end{equation}
Examining the orbifold expansion in terms of the algebraic coordinate $\psi=z^{-1/3}$ and using the knowledge (\ref{asympofABC}) of the analytic continuation of $\theta t=\omega_{0}\sim \psi$, we can check that $F^{(1)}$ has no logarithmic singularity in $\psi$ and hence no particles of an effective theory become massless at this point.
We can furthermore compute
\begin{equation} \label{genus1P2}
\partial_t \,F^{(1)}=-\frac{1}{6} (1+C_0) G_1^{-3} \, T_2\,,
\end{equation}
giving $f_z^1=0$ in Eq.~(\ref{genus1mod}).

\subsubsection{Higher genus amplitudes and duality}
The anomaly recursion in terms of the generators (\ref{specialrec1}) becomes:
\begin{equation}
\frac{\partial F^{(g)}}{\partial T_2}=\frac{1}{2} \sum_{h=1}^{g-1} \partial_t F^{(h)}\, \partial_t F^{(g-h)}+\frac{1}{2} \partial_t\partial_t F^{(g-1)}\,.
\end{equation}
Together with the initial amplitude (\ref{genus1P2}) the higher genus amplitudes can be obtained up the the addition of the holomorphic ambiguity which only needs to take into account the singularity at the conifold expansion locus and is of the form:
\begin{equation}
A^{(g)}=\sum_{i=0}^{2g-2} a_i\, C_0^i\,,
\end{equation}
the $2g-1$ coefficients $a_i$ can be fixed by first using the duality discussed in Section~\ref{sec:duality} and then implementing the vanishing of the subleading singularities in terms of the right flat coordinate at the conifold. This has been done using methods of analytic continuation in Refs.~\cite{Haghighat:2008gw,Alim:2008kp}, the use of the duality operation simplifies this considerably. A simple counting then shows that this model can be recursively solved.

For example $F^{(2)}$ can be computed to be:
  \begin{equation}
 F^{(2)}= \frac{ (1+C_0)^2 \, T_2 \left( 3\,G_1^4-9G_1^2\, T_2 + 10 T_2^2\right)}{432 \,G_1^6} + a_0 +a_1 C_0+ a_2\, C_0^2\,,
 \end{equation}
the coefficients $a_0,a_1$ and $a_2$ will be fixed using the duality in the following.

\subsubsection*{Duality action and conifold expansion}
The Fricke involution (\ref{fricke}) on this choice of generators becomes\footnote{Strictly speaking it is the non-holomorphic completion of $\widehat{T}_{2}={\widehat{E}\over 2}$ transforms this way.}:
\begin{equation}
T_2\rightarrow 3 \tau^2 \,T_2\,, \quad G_1 \rightarrow -i \sqrt{3}\, \tau\,G_1\,, \quad C_0 \rightarrow \frac{1}{C_0}\,.
\end{equation}
The relevant flat coordinate at the conifold corresponds to the analytic continuation of the period $F_t$, we introduce the following normalization for convenience:
\begin{equation}
t_c=\frac{1}{i 3\sqrt{3}} F_t\,,
\end{equation}
The duality operation will give the expansion of $F^{(g)}$ in terms of the modular coordinate $\tau$ centered at the new cusp. To obtain the expansion in terms of $F_t$ we look at the following relation:
\begin{equation}
\partial_\tau F_t= -\frac{1}{3} \tau \frac{G_1^3}{1+C_0}\,,
\end{equation}
 which follows from the definitions and becomes after the duality transformation:
 \begin{equation}
 \partial_{\tau} t_c=\frac{1}{i 3\sqrt{3}} \partial_\tau\, F_t =\frac{1}{27} \frac{C_0 G_1^3}{1+C_0}\,,
 \end{equation}
this latter expression admits a Fourier expansion in $q_\tau=\exp(2\pi
i \tau)$, which can be integrated and inverted to give:
 \begin{equation}
q_{\tau}(t_c) =t_c-\frac{3 t_c^2}{2}+\frac{3 t_c^3}{2}+\frac{19 t_c^4}{8}+\mathcal{O}(t_c^5)\,.
\end{equation}
In terms of $t_c$, the singularity in $F^{(g)}(t_c)$ is expected not to have subleading terms. This together with the contribution from constant maps can be used to fix the higher genus amplitudes. For example at genus 2 we find the answer:
\begin{equation}
F^{(2)}=\frac{ (1+C_0)^2 \, T_2 \left( 3\,G_1^4-9G_1^2\, T_2 + 10 T_2^2\right)}{432 \,G_1^6} -\frac{11}{17280}-\frac{7}{4320}C_0-\frac{1}{1080} C_0^2\,.
\end{equation}
Applying the duality transformation we find:

\begin{equation}
F^{(2)}_D= -\frac{\left(11 C_0^2+28 C_0+16\right) C_1^6+120 (C_0+1)^2 C_1^4 T_2+360
   (C_0+1)^2 C_1^2 T_2^2+400 (C_0+1)^2 T_2^3}{17280 C_0^2 C_1^6}\,,
\end{equation}
with the expansion
\begin{equation}
F^{(2)}_D(t_c)=-\frac{1}{58320 t_c^2}+\frac{1}{51840}+\frac{t_c}{720}-\frac{3187 t_c^2}{518400}+\frac{239
   t_c^3}{12960}-\frac{19151 t_c^4}{483840}+\mathcal{O}(t_c^5)\,.
      \end{equation}
Higher genus amplitudes can be obtained easily by this procedure,
since those are rather lengthy we will only display genus 3 results
in Appendix \ref{highergenus}.


\subsection{Local $E_n$ geometries}

In the following we shall consider the mirror manifolds (B-model) of
the canonical bundles of $\mathbb{P}^{2}$ and del Pezzo surfaces
$dP_{n},\, n=5,6,7,8$ (A-model). These are non-compact CY
threefolds, given by a set of polynomial equations $W_{E_{n}}=0$ in
certain weighted projective spaces in Ref.~\cite{Lerche:1996ni}. For
these B-model geometries, as shown in e.g.
Ref.~\cite{Lerche:1996ni}, the Picard--Fuchs equations factor in the
way described in (\ref{factorizepf}). The corresponding families of
elliptic curves are called of $E_{n}$ type in the literature, see
e.g. Refs. \cite{Lian:1994zv,Klemm:1996}, and will be recalled
below. In this work, we shall refer to the B-model non-compact CY
threefolds as local $E_{n}$ del Pezzo. Mirror symmetry for these
geometries and some related discussions can be found e.g. in Refs.
\cite{Lian:1994zv,Klemm:1996,Lerche:1996ni,Chiang:1999tz,Klemm:1999gm,
Katz:1999xq,Mohri:2001zz}.

\subsubsection{Initial data}\label{initialdata}
\subsubsection*{\it Families of non-compact CY threefolds}

The non-compact CY threefold families are defined by the
following charge vectors, see for example Ref.~\cite{Lerche:1996ni}.
These charge vectors correspond to the
Mori generators on the A-model geometry and give the PF operators of the B-model geometry:
\begin{equation}
  \label{eq:Mori}
  \begin{aligned}
 N=1^*:\quad    l^{(1)} & = (-6;3,2,1,-1,1)\,,\\
 N=2:\quad      l^{(2)} & = (-4;2,1,1,-1,1)\,.\\
 N=3:\quad     l^{(3)} & = (-3;1,1,1,-1,1)\,,\\
 N=4:\quad     l^{(4)} & = (-2,-2;1,1,1,1,-1,1)\,.
  \end{aligned}
\end{equation}
We recall $$d_N = 432, 64, 27, 16,\quad \text{
for } N = 1^*,2,3,4,$$ respectively. From the Mori generators we
obtain the following Picard--Fuchs operators:
\begin{equation}
  \label{eq:PFdelPezzo}
  \cL^{(N)}_{dP} = \cL_{ell}^{(N)}\circ \,\theta\,,
\end{equation}
with $\cL_{ell}^{(N)}$ being the Picard--Fuchs operators for
elliptic curves as shown in (\ref{deflc})\footnote{Here the coordinate $z$ is chosen to be the same as $z_{c}$.}:
\begin{equation}
  \label{eq:PFops}
    \cL_{ell}^{(N)} = \theta^2 - d_N \,z (\theta + 1/r) (\theta + 1-1/r)\,,
\end{equation}
with $r=6,4,3,2$ for $N=1^*,2,3,4$. These operators have regular singularities at $z=0$,
$z=\frac{1}{d_N}$ and $z=\infty$ on the moduli space, corresponding to the large complex structure limit, conifold and orbifold point, respectively.  A pair of linearly
independent solutions near $z=0$ is given by the hypergeometric
functions (\ref{hypersols})
\begin{equation}
  \label{eq:3}
  \omega_{0}={}_2F_{1}(1/r,1-1/r;\,1;\, d_{N}z)\,,\quad  \omega_{1}={i \over \sqrt{N}}~{}_2F_{1}(1/r,1-1/r;\,1;\, 1-d_{N}z)\,.
\end{equation}
The normalization is chosen so that $\tau={\omega_{1}\over
\omega_{0}}$. The monodromies about $z=0, \frac{1}{d_N},
\infty$ are
\begin{equation}
  \label{eq:monodromies}
  M_0 =
  \begin{pmatrix}
    1 & 1\\ 0 & 1
  \end{pmatrix}
  = T\,,
  M_1 =
  \begin{pmatrix}
    1 & 0 \\ -N & 1\\
  \end{pmatrix}
  = -S T^{N} S,
  M_\infty = M_1M_0\,,
\end{equation}
respectively, where $S$ and $T$ are the standard generators of
$\overline{\Gamma}(1)=SL(2,\mZ)$. Hence the monodromy group is
$\Gamma_0(N)$ if $N\neq 1$, and coincides with the modular group. For
$N=1^*$, the monodromy group is $\Gamma(1)$. However, there
is an additional $\mZ_2$ symmetry $z \mapsto 1 - d_1z$ under which the
corresponding elliptic curves are isomorphic. This can be shown e.g. using
the analytic continuation formulae for ${}_2F_1$. As a consequence,
the modular group is actually the subgroup of index 2 in $\Gamma(1)$
generated by $T^2$ and $T^{-1}S$, denoted by $\Gamma_0(1)^*$.

\subsubsection*{\it Families of the elliptic curves of $E_{n}$ type}\label{Encurvefamilies}

Now we are going to explain more details about the families of
elliptic curves of $E_n$ type mentioned at the beginning of the section.
The explicit equations, $j$ invariants, as well as the
Picard-Fuchs operators are summarized as follows, see
Refs.~\cite{Lian:1994zv,Lerche:1996ni,Klemm:1996} for more details.

\begin{equation}
E_{5}:\, \left\{
\begin{array}{c l}
x_{1}^{2}+x_{3}^{2}-z^{-{1\over 4}} x_{2}x_{4}&=0\\
x_{2}^{2}+x_{4}^{2}-z^{-{1\over 4}} x_{1}x_{3}&=0
\end{array}\right.
\quad
j(z)={(1+224z+256z^{2})^{3}\over z(1-16z)^{4}}\,,
\quad
\mathcal{L}=\theta^{2}-4z(2\theta+1)^{2}.
\end{equation}
The base of this family of elliptic curves is the modular curve $X_{0}(4)$. It has
three singular points: two cusp classes $[i\infty],[0]$ corresponding to $z=0,1/16$ respectively;
and the cusp class $[1/2]$ corresponding to $z=\infty$.

\begin{equation}
E_{6}:\,x_{1}^{3}+x_{2}^{3}+x_{3}^{3}-z^{-{1\over 3}} x_{1}x_{2}x_{3}=0 \,,\quad
j(z)={(1+216z)^{3}\over z(1-27z)^{3}}\,,\quad
\mathcal{L}=\theta^{2}-3z(3\theta+1)(3\theta+2)\,.
\end{equation}
The base of this family of elliptic curves is the modular curve $X_{0}(3)$. It has
three singular points: two cusp classes $[i\infty],[0]$ corresponding to $z=0,1/27$ respectively;
and the cubic elliptic point $[\rho]$ corresponding to $z=\infty$, where $\rho=\exp (2\pi i/3)$.

\begin{equation}
E_{7}:\,x_{1}^{4}+x_{2}^{4}+x_{3}^{2}-z^{-{1\over 4}} x_{1}x_{2}x_{3}=0 \,,\quad
j(z)={(1+192z)^{3}\over z(1-64z)^{3}}\,,\quad
\mathcal{L}=\theta^{2}-4z(4\theta+1)(4\theta+3)\,.
\end{equation}
The base of this family of elliptic curves is the modular curve $X_{0}(2)$. It has
three singular points: two cusp classes $[i\infty],[0]$ corresponding to $z=0,1/64$ respectively;
and the quadratic elliptic point $[i]$ corresponding to $z=\infty$.

\begin{equation}
E_{8}:\,x_{1}^{6}+x_{2}^{3}+x_{3}^{2}-z^{-{1\over 6}} x_{1}x_{2}x_{3}=0\,, \quad
j(z)={1\over z(1-432z)}\,,\quad
\mathcal{L}=\theta^{2}-12z(6\theta+1)(6\theta+5)\,.
\end{equation}
The base of this family of elliptic curves is the curve $X_{0}(1)^{*}$. It has
three singular points: two cusp classes $[i\infty],[0]$ corresponding to $z=0,1/432$ respectively;
and the cubic elliptic point $[\rho]$ corresponding to $z=\infty$.

\subsubsection*{\it Genus 0: Yukawa couplings}

In the above we have identified the moduli spaces of the local
$E_{n}$ del Pezzo geometries with the bases of the corresponding
$E_{n}$ curve families and thus the modular curves $X_{0}(N)$, with
$N=4,3,2,1^{*}$ when $n=5,6,7,8\,,$ respectively. From the
Picard-Fuchs equation of the $E_{n}$ elliptic curve families, we get
\begin{equation}
C_{z}=-\int_{E_{z}}\omega_{z}\wedge \partial_{z}\omega_{z}= {1\over
z(1-d_{N}z)}\,,
\end{equation}
where $\omega_{z}$ is the holomorphic one form on the elliptic curve $E_{z}$.
From this and the following equation
\begin{equation}\label{curveyukawa}
C_{2\pi i\tau}={1\over 2\pi i}{dz\over d\tau} C_{z}{1\over
A^{2}}=1\,,
\end{equation}
we get (\ref{gaussschwarz}). The above equation (\ref{curveyukawa}),
see e.g. Refs.~\cite{Lian:1994zv, Zagier:1998, Hosono:2008ve},
represents the fact that there is no quantum correction to the
Yukawa coupling of elliptic curves. It follows that the Yukawa
coupling for the corresponding local $E_{n}$ del Pezzo is given by
\begin{equation}
 C_{zzz}={\kappa\over z^{3}(1-d_{N}z)}\,,
\end{equation}
where $\kappa$ is the classical triple intersection on the A-model CY geometry, as described in (\ref{yukawa}).
Therefore, we have
\begin{equation}
 C_{ttt}={\kappa\over (\theta t)^{3}(1-d_{N}z)}\,.
\end{equation}

\subsubsection*{\it Genus 1}

The topological invariants for the corresponding A-model non-compact
CY threefolds can be found in \cite{Lerche:1996ni,
Chiang:1999tz}\footnote{The Chern number $c_{2}$ is the integral of
the Chern class previously denoted by $\int c_{2}J$ in
(\ref{bdyconditionatlcsl}).}
\begin{equation*}
\kappa=n-9\,, \quad c_{2}=-12+2(9-n)\,, \quad  c_{3}=\chi =-2h(E_{n})\,
, \quad h(E_{n})=8,12,18, 30\,.
\end{equation*}
Near the large complex structure limit, the genus one amplitude, denoted by $F_{lcs}^{(1)}\,$, is given by
\begin{equation}
F_{lcs}^{(1)}=-
{1\over 2}\log \theta t+\log (1-\alpha)^{a}\alpha^{b }\,.
\end{equation}
The constant $a$ is universal and is given by $a=-{1\over 12}\,$, while $b=-{c_{2}\over 24}\,$.
Now we will compute the singular behavior of $F^{(1)}_{orb}$, this is the analytic continuation of $F_{lcs}^{(1)}$ to the orbifold point.
In each case above, according to (\ref{asympofABC}) we have near the
orbifold point $\alpha=\infty$,
\begin{equation}
\theta t =\omega_{0}\sim \alpha^{-{1\over r}}(1+\mathcal{O}(\alpha^{-{1\over r}}))\,.
\end{equation}
Hence
\begin{equation}
F_{orb}^{(1)}
\sim -{1\over 12}\log (\alpha^{-{1\over r}})^{6} (1+\mathcal{O}(\alpha^{-{1\over r}}))(1-\alpha)\alpha^{c_{2}\over 2}
\sim -{1\over 12}\log \alpha^{-{6\over r}+1+{c_{2} \over 2}}\,.
\end{equation}
Changing to the local coordinate $\psi=\alpha^{-{1\over r}}$ near the
orbifold point, we then have
\begin{equation}
F_{orb}^{(1)}\sim -{1\over 12}\log\psi^{6-r(1+{c_{2}\over 2})}= -{1\over 12}\log\psi^{h(E_{n})}\,.
\end{equation}
The numbers $h(E_{n})=6-r(1+{c_{2}\over 2})$ for $n=5,6,7,8$ cases are given by
$8,12,18,30$, respectively, they are the dual Coxeter numbers \cite{Chiang:1999tz, Lerche:1996ni}
of the Lie algebra $E_{n}$. Due to the singular behavior of the genus one amplitude, the higher genus
amplitudes will be singular from the polynomial recursion obtained from holomorphic anomaly equations. This higher genus singularity appears implicitly in the ambiguities determined in Ref.~\cite{Katz:1999xq}. The expansions of the higher genus amplitudes which we obtain in terms of the flat coordinates in this region exhibit singular behavior, the systematics of which will be discussed elsewhere.


\subsubsection{Special polynomial ring as ring of quasi modular forms}

Similar to the local $\mathbb{P}^{2}$ case, for these local $E_{n}$ del Pezzo
geometries, the special polynomial ring (\ref{relspecial}) constructed out of special
geometry will reduce to exactly the ring of quasi modular forms (\ref{quasiring})
constructed using the arithmetic of the modular curves with suitable choices of holomorphic functions.

Recall that for local $E_{n}$ geometries, we have the following table:
\begin{equation}
\begin{array}{c|cccc}
n&5&6&7&8\\
N&4&3&2&1^{*}\\
\kappa&-4&-3&-2&-1\\
d_{N}&16&27&64&432\\
r&2&3&4&6
\end{array}
\end{equation}
In terms of the quasi modular form generators in (\ref{quasiring}),
we have the following expressions for Yukawa coupling and genus one
amplitude:
\begin{eqnarray}
C_{ttt}&=&{\kappa \over A^{3} (1-\alpha)}={\kappa A^{r-3}\over B^{r}}\,,\\
F^{(1)}&=&-{1\over 2}\log A+\log \left(B^{-{r\over 12}} C^{-{rc_{2}\over 24}}/A^{{-{r\over 12}}-{rc_{2}\over 24}}\right)\,,\\
\partial_{\tau}F^{(1)}&=&-{1\over 4r}E+ {1\over
2}A^{2} \left({1\over 12}-{c_{2}\over 24}-({1\over 12}+{c_{2}\over
24})(C^{r}-B^{r})A^{-r}\right)\,.
\end{eqnarray}
We make the following choices for the ambiguities in (\ref{relspecial}):
\begin{equation}
zs_{zz}^{z}+1={1\over 2r}{\alpha-\beta\over \beta}\,,\quad
\tilde{h}_{zz}^{z}=\frac{1}{(2r)^2 \beta ^2}\,,\quad  \tilde{h}^{z}_{zzz}={\alpha\over \beta}-{3\over 2r}{\alpha-\beta\over \beta}\,,
\end{equation}
where $\alpha=d_{N} z$ and $\beta=\Delta=1-\alpha$ as before.
It follows then
\begin{eqnarray}
T_{2}=-{1\over 2r\kappa}E={1\over 2rN}E\,,\quad C_{0}={C^{r}\over B^r}\,,\quad G_{1}=A\,,\quad K_{0}={B^{r}\over A^{r}}=(1+C_{0})^{-1}\,,
\end{eqnarray}
The other generators vanish in the holomorphic limit. It is immediate to see that for these special choices of ambiguities, the special polynomial ring (\ref{relspecial})
is equivalent to the ring of quasi modular forms (\ref{quasiring}) for each of these geometries.
Below we briefly summarize the results.
\subsubsection*{\it Local $E_{5}$ del Pezzo}
Holomorphic ambiguities:
\begin{eqnarray}
s_{zz}^{z}=-\frac{3}{2}+\frac{1}{4 \Delta }\,,\quad
\tilde{h}_{zz}^{z}={1\over 16\Delta^2}\,, \quad
\tilde{h}^{z}_{zzz}=\frac{1 }{2  }+\frac{1}{4\Delta}\,.
\end{eqnarray}
Special polynomial ring generators:
\begin{eqnarray}
T_{2}={1\over 16}E\,,\quad C_{0}={C^{2}\over B^2}\,,\quad G_{1}=A,\quad K_{0}={B^{2}\over A^{2}}=(1+C_{0})^{-1}\,.
\end{eqnarray}
Differential ring structure:
\begin{align}
&\partial_{\tau} C_0=G_{1}^{2}C_{0}\,,\\
& \partial_{\tau} G_1=\frac{1}{4}\left(16G_1\, T_2+G_1^3\left(\frac{C_0-1}{C_0+1}\right)\right)\,,\\
& \partial_{\tau} T_2=\frac{1}{4}(16 T_2^2-\frac{G_1^4}{16})\,.
\end{align}
\subsubsection*{\it Local $E_{6}$ del Pezzo}
Holomorphic ambiguities:
\begin{eqnarray}
s_{zz}^{z}=-\frac{4}{3}+\frac{1}{6 \Delta }\,,\quad
\tilde{h}_{zz}^{z}={1\over 36\Delta^2 }\,, \quad
\tilde{h}^{z}_{zzz}=\frac{1}{2 \Delta }\,.
\end{eqnarray}
Special polynomial ring generators:
\begin{eqnarray}
T_{2}={1\over 18}E\,,\quad C_{0}={C^{3}\over B^3}\,,\quad G_{1}=A\,,\quad K_{0}={B^{3}\over A^{3}}=(1+C_{0})^{-1}\,.
\end{eqnarray}
Differential ring structure:
\begin{align}
&\partial_{\tau} C_0=G_{1}^{2}C_{0}\,,\\
& \partial_{\tau} G_1=\frac{1}{6}\left(18G_1\, T_2+G_1^3\left(\frac{C_0-1}{C_0+1}\right)\right),\,\\
& \partial_{\tau} T_2=\frac{1}{6}\left(18 T_2^2-\frac{G_1^4}{18}\right)\,.
\end{align}
\subsubsection*{\it Local $E_{7}$ del Pezzo}
Holomorphic ambiguities:
\begin{eqnarray}
s_{zz}^{z}=-\frac{5}{4}+\frac{1}{8 \Delta }\,,\quad
\tilde{h}_{zz}^{z}={1\over 64\Delta^2}\,, \quad
\tilde{h}^{z}_{zzz}=-{1\over 4}+{5\over 8\Delta}\,.
\end{eqnarray}
Special polynomial ring generators:
\begin{eqnarray}
T_{2}={1\over 16}E\,,\quad C_{0}={C^{4}\over B^4}\,,\quad G_{1}=A\,,\quad K_{0}={B^{4}\over A^{4}}=(1+C_{0})^{-1}\,.
\end{eqnarray}
Differential ring structure:
\begin{eqnarray}
&&\partial_{\tau} C_0=G_{1}^{2}\,C_{0}\,,\\
&& \partial_{\tau} G_1=\frac{1}{8}\left(16G_1\, T_2+G_1^3\left(\frac{C_0-1}{C_0+1}\right)\right)\,,\\
&& \partial_{\tau} T_2=\frac{1}{8}(16 T_2^2-\frac{G_1^4}{16})\,.
\end{eqnarray}
\subsubsection*{\it Local $E_{8}$ del Pezzo}
Holomorphic ambiguities:
\begin{eqnarray}
s_{zz}^{z}=-\frac{7}{6}+\frac{1}{12 \Delta }\,,\quad
\tilde{h}_{zz}^{z}=\frac{1}{144 \Delta ^2}\,, \quad
\tilde{h}^{z}_{zzz}=-{1\over 2}+\frac{3}{4 \Delta }\,.
\end{eqnarray}
Special polynomial ring generators:
\begin{eqnarray}
T_{2}={1\over 12}E\,,\quad C_{0}={C^{6}\over B^6}\,,\quad G_{1}=A\,,\quad K_{0}={B^{6}\over A^{6}}=(1+C_{0})^{-1}\,.
\end{eqnarray}
Differential ring structure:
\begin{eqnarray}
&&\partial_{\tau} C_0=G_{1}^{2}\,C_{0}\,,\\
&& \partial_{\tau} G_1=\frac{1}{12}\left(12G_1\, T_2+G_1^3\left(\frac{C_0-1}{C_0+1}\right)\right)\,,\\
&& \partial_{\tau} T_2=\frac{1}{12}(12 T_2^2-\frac{G_1^4}{12})\,.
\end{eqnarray}

\subsubsection{Higher genus amplitudes and duality}
Similar to what we did for local $\mathbb{P}^{2}$ case, after we have identified the moduli spaces with modular curves,
 we could use the polynomial recursion,
combing the Fricke involution and boundary conditions
(\ref{bdyconditionatlcsl}), (\ref{Gap}) to fix the ambiguities in
$F^{(g)}$ for the local $E_{n}$ del Pezzo geometries.  As in the local
$\mathbb{P}^{2}$ case, when considering the holomorphic limits of
topological string amplitudes, effectively the Fricke involution
(\ref{fricke}) acts on these generators according to
\begin{equation}
T_2\rightarrow N \tau^2 \,T_2\,, \quad G_1 \rightarrow {\sqrt{N}\over i}\, \tau\,G_1\,, \quad C_0 \rightarrow \frac{1}{C_0}\,.
\end{equation}
The same procedure determines the topological string amplitudes.  In
the following, the genus 2 expressions are given including the
expansions in terms of the vanishing period $t_{c}$ near the
conifold point $\Delta=0$, genus 3 expressions can be found in
Appendix \ref{highergenus}.

\subsubsection*{\it Local $E_{5}$ del Pezzo}
\begin{eqnarray*}
&F^{(2)}=-\frac{275+73 C_{0}^2+C_{0}^3+16 \chi+C_{0} (119+16 \chi)}{92160 (1+C_{0})}+
\frac{\left(109-50 C_{0}+C_{0}^2\right) T_2}{1152 G_{1}^2}
-\frac{\left(23+24 C_{0}+C_{0}^2\right) T_2^2}{24 G_{1}^4}+\frac{10 (1+C_{0})^2 T_2^3}{3 G_{1}^6}\,,
\end{eqnarray*}
\begin{eqnarray*}
&F^{(2)}_{D}(t_{c})=-\frac{1}{960 {t_c}^2}+\frac{-35-\chi}{5760}+\frac{1733 {t_c}}{491520}-\frac{99421 {t_c}^2}{58982400}+\frac{3349 {t_c}^3}{4718592}-\frac{10556017 {t_c}^4}{38050725888}+\mathcal{O}(t_{c}^{5})\,.
\end{eqnarray*}
\subsubsection*{\it Local $E_{6}$ del Pezzo}
\begin{eqnarray*}
&F^{(2)}=\frac{-500-28 C_{0}-16 C_{0}^2-27 \chi}{155520}+
\frac{\left(13-2 C_{0}+C_{0}^2\right) T_2}
{144 G_{1}^2}-\frac{\left(11+14 C_{0}
+3 C_{0}^2\right) T_2^2}{16 G_{1}^4}
+\frac{15 (1+C_{0})^2 T_2^3}{8 G_{1}^6}\,,
\end{eqnarray*}
\begin{eqnarray*}
&F^{(2)}_{D}(t_{c})=-\frac{1}{720 {t_c}^2}+\frac{-3866-81 \chi}{466560}+\frac{641 {t_c}}{174960}-\frac{5129587 {t_c}^2}{3401222400}+\frac{1287599 {t_c}^3}{2295825120}-\frac{451320911 {t_c}^4}{2314191720960}+\mathcal{O}(t_{c}^{5})\,.
\end{eqnarray*}
\subsubsection*{\it Local $E_{7}$ del Pezzo}
\begin{eqnarray*}
&F^{(2)}=\frac{-1835+36 C_{0}-133 C_{0}^2-64 \chi}{368640}+
\frac{\left(493+14 C_{0}+61 C_{0}^2\right) T_2}{4608 G_{1}^2}-\frac{\left(47+64 C_{0}
+17 C_{0}^2\right) T_2^2}{96 G_1^4}+\frac{5 (1+C_{0})^2 T_2^3}{6 G_{1}^6}\,,
\end{eqnarray*}
\begin{eqnarray*}
&F^{(2)}_{D}(t_{c})=
-\frac{1}{480 {t_c}^2}+\frac{-2173-32 \chi}{184320}+\frac{18777 {t_c}}{5242880}-\frac{8668429 {t_c}^2}{7549747200}+\frac{1382167 {t_c}^3}{4026531840}-\frac{37720723573 {t_c}^4}{389639433093120}+\mathcal{O}(t_{c}^{5})\,.
\end{eqnarray*}
\subsubsection*{\it Local $E_{8}$ del Pezzo}
\begin{eqnarray*}
&F^{(2)}=\frac{-1825+36 C_{0}-299 C_{0}^2-36 \chi}{207360}
+\frac{5 \left(29+6 C_{0}+5 C_{0}^2\right) T_2}{1152 G_{1}^2}-\frac{\left(25+36 C_{0}+11 C_{0}^2\right) T_2^2}
{96 G_{1}^4}+\frac{5 (1+C_{0})^2 T_2^3}{24 G_{1}^6}\,,
\end{eqnarray*}
\begin{eqnarray*}
&F^{(2)}_{D}(t_{c})
=-\frac{1}{240 {t_c}^2}+\frac{-3011-27 \chi}{155520}+\frac{21151 {t_c}}{5971968}-\frac{223349623 {t_c}^2}{290237644800}+\frac{101569651 {t_c}^3}{626913312768}-\frac{820546280317 {t_c}^4}{25277144770805760}+\mathcal{O}(t_{c}^{5})\,.
\end{eqnarray*}

In the above, we have normalized so that the vanishing period
$t_{c}(\Delta)$ has the form
$t_{c}(\Delta)=\Delta+\mathcal{O}(\Delta^{2})$ near the conifold
point $\Delta=0$. From these expansions one can see that for each of
these local $E_n$ del Pezzo geometries, at genus $2$, the gap condition
takes the form $F_{D}^{(2)}(t_{c})={1\over 240\kappa
t_{c}^{2}}+\mathcal{O}(t_{c}^{0})$.

As a consistency check, we have checked that all of
these modular functions reproduce the integral Gopakumar-Vafa
invariants listed in \cite{Katz:1999xq}.


\subsection{Compact geometry}
It was shown in the previous examples that the special differential polynomial ring defined in (\ref{relspecial})
gives the rings of quasi modular forms for non-compact geometries with a duality group for which these are known. In the following we will define the analogous differential ring in terms of the analogous coordinates for the example of the quintic, a compact CY threefold. Mirror symmetry for the quintic is the classical example, studied in detail in Ref.~\cite{Candelas:1990rm}. The polynomial structure of higher genus topological string amplitudes for the quintic was put forward in Ref.~\cite{Yamaguchi:2004bt} and used in Ref.~\cite{Huang:2006hq} together with the boundary conditions to enhance higher genus computations. The generalization of the polynomial construction \cite{Alim:2007qj} gives slightly different generators. The freedom of adding holomorphic functions to the generators was discussed in Refs.~\cite{Alim:2008kp,Hosono:2008ve} and used in \cite{Hosono:2008ve} to discuss the rationality of the holomorphic functions appearing in the polynomial setup. A ring of functions for the quintic as a generalization of the Eisenstein series was proposed in Ref.~\cite{Movasati:2011zz}. The discussion in this work suggests that it is the same ring.

\subsubsection{Special polynomial ring}
A discussion of mirror symmetry for the quintic can be found in Refs.~\cite{Candelas:1990rm,Ceresole:1993qq}. The Yukawa coupling is given by \footnote{For the mirror quintic in terms of an algebraic coordinate on the moduli space, and adopting the convention of multiplying lower tensorial indices by $z$.}
\begin{equation}
C_{zzz}=\frac{5}{\Delta}\,,\quad \Delta=(1-3125 z)\,.
\end{equation}
We fix the holomorphic functions appearing in (\ref{relspecial}) as in Refs.~\cite{Hosono:2008ve,Alim:2012gq}:
\begin{equation}
s_{zz}^z=-\frac{8}{5}\, \quad \tilde{h}_{zz}^{z}=\frac{1}{5\Delta}\,,\quad \tilde{h}_{zz}=-\frac{1}{25\Delta}\,,\quad \tilde{h}_z=\frac{2}{625 \Delta}\,,\quad k_{zz}=\frac{2}{25}\,.
\end{equation}
The generator of rational functions (\ref{ratgen}) becomes:
\begin{equation}
C_0=\frac{3125 z}{1-3125z}\,,
\end{equation}
giving the special ring relations:
\begin{eqnarray}\label{relspecialquintic}
\partial_{\tau}C_0 &=&C_0\,(1+C_0)\,K_0\,G_1^2\,, \\
\partial_{\tau} K_0 &=&-2K_0\,K_2- C_0 K_0^2\, G_1^2\,,\\
\partial_{\tau} G_1&=& 2G_1\,K_2-5  G_1\,T_2\, -\frac{3}{5} K_0 G_1^3\,,\\
\partial_{\tau} K_2&=&3K_2^2-15 K_2\,T_2-25 T_4+ \frac{2}{25}\,K_0^2\,G_1^4 - \left(\frac{9}{5}+C_0\right)K_0\,G_1^2\,K_2\,,\\
\partial_{\tau} T_2&=&2K_2\,T_2-5 T_2^2+10 T_4+\frac{1}{25}(1+C_0)\,K_0^2 G_1^4\,,\\
\partial_{\tau} T_4&=&4 K_2 T_4-15 T_2\,T_4+ 10 T_6- \left(\frac{9}{5}+C_0\right) K_0\, G_1^2 \, T_4 -\frac{2}{125} K_0^2\, G_1^4 \,T_2\nonumber\\
&&-\frac{1}{625}(1+C_0) K_0^3\, G_1^6 \,,\\
\partial_{\tau} T_6&=& 6 K_2\, T_6-30 T_2 \,T_6+\frac{5}{2} T_4^2-\frac{2}{125} K_0^2\, G_1^4 \,T_4+\frac{2}{78125}(1+C_0) K_0^4\, G_1^8\nonumber\\
 &&-2 \left(\frac{9}{5}+C_0\right)\, K_0\,G_1^2\,T_6 \,.
\end{eqnarray}

\subsubsection{Higher genus amplitudes}
The higher genus amplitudes can be obtained from Eq.~(\ref{specialrec1}), starting with the initial data of genus $1$ which can be fixed using the topological data needed (\ref{genus1}) \footnote{See Ref.~\cite{Alim:2012gq} and references therein.}:
\begin{equation}
n=1\,,\quad \chi=-200\,,\quad  s=-\frac{31}{12}\,, \quad r=-\frac{1}{12}\,,
\end{equation}
which gives in terms of the generators
\begin{equation}
F^{(1)}=-\frac{1}{12} \log\left(3125^{-25} (1+C_0)^{36} C_0^{25}\,K_0^{62}\,G_1^6 \right)\,,
\end{equation}
leading to the initial correlation function
\begin{equation}
F^{(1)}_t= \frac{(5 C_0-107) G_1^2 K_0+560 K_2+150
   T_2}{60 G_1^3 K_0}\,.
\end{equation}
Using the boundary conditions, the genus two amplitude for example can be determined to be:
\begin{eqnarray}
F^{(2)}&=&\frac{25 \left(65 C_0^2-46 C_0+2129\right) G_1^4
   K_0^2 T_2+\left(30 C_0^3+113 C_0^2-488
   C_0-571\right) G_1^6 K_0^3}{36000 \, K_0^2\,G_1^6}\\ &&+\frac{5000 G_1^2
   K_0 \left(6 (C_0-4) T_2^2+5 (107-5 C_0)
   T_4\right)+62500 \left(3 T_2^3-60 T_2
   T_4+1120 T_6\right)}{36000\,K_0^2\,G_1^6}\,. \nonumber
\end{eqnarray}
We have thus shown in a compact example the general properties of the special polynomial ring which we defined and the expressions of higher genus amplitudes in terms of these generators. The study of the analog of the duality action and a more careful analysis of the arithmetics of the moduli space parameterized in terms of $\tau$ will be addressed in the future.

\section{Conclusions}
We studied differential rings of polynomial generators, defined using the special geometry of the moduli space of a CY geometry and in terms of which the topological string amplitudes with this CY as a target manifold can be expressed. The polynomial generators we use are those defined in Ref.~\cite{Alim:2007qj} as a generalization of those in Ref.~\cite{Yamaguchi:2004bt}. The definitions in Ref.~\cite{Alim:2007qj} only use the special geometry as an input and can thus be used for any CY target. We defined new generators based on these and made a special choice of coordinate $\tau$ on the moduli space. These constructions allowed us to assign a new grading to the differential ring of generators such that the derivative w.r.t. $\tau$ strictly increases the grading by $2$ and such that the functions obtained from the topological string amplitudes with $n$ insertions $F^{(g)}_{n}$ have degree $-n$.

In a number of examples we showed that the special ring which we defined in generality coincides with the ring of quasi modular forms where the latter are known. The grading becomes the modular weight and using the polynomial form the BCOV anomaly we can construct $F^{(g)}$ as quasi modular functions for these examples. We studied the action of the Fricke involution on the quasi modular forms and showed that on the level of the topological string amplitudes it exchanges the large complex structure and the conifold expansion loci, giving a form of an electric-magnetic duality.

Motivated by the results relating the special ring of generators to the quasi modular forms in the known examples we constructed the analogous ring for the example of the quintic. Since the construction of the special ring only relied on manipulations of the special geometry this paves the way to exploring the analogs of quasi modular forms for many more examples. For the moment we only provided the analogous differential ring of functions awaiting a more thorough study of the duality group of the quintic and duality groups of other compact target spaces.

The exchange of cusps of the moduli space for Riemann surfaces as a geometric origin of $\mathcal{N}=2$ dualities has been explored in much more detail in Refs.~\cite{Argyres:2007cn,Gaiotto:2009we}, it would be interesting to attach generating functions of BPS degeneracies to a given family of theories and express these in terms of quasi modular forms, the action of the Fricke involution on these should encode non-trivial wall crossing phenomena.

For general compact CY threefolds, it is challenging to identify the moduli space of complex structures $\mathcal{M}$
with certain modular curves. This is basically due to the lack of a good understanding of the global Torelli theorem which asserts that
the period map is an one to one map from the moduli space of complex structures to the period domain which carries
modular group actions. We hope to explore more of the arithmetic
properties of the special geometry ring for compact CY threefolds in the future.

The use of the coordinate $\tau$ on the moduli space was motivated to establish the relation to known modularity in non-compact examples \cite{Aganagic:2006wq}. The non-trivial map between the exponentiated K\"ahler moduli $q_t=\exp(2\pi i t)$ and $q_\tau=\exp (2 \pi i \tau)$ begs for an explanation of its enumerative and physical content, so do the $q_\tau$ expansions of the topological string amplitudes.
$\tau$ in the non-compact cases corresponds to the flat coordinated on a lower dimensional geometry, the mirror curve, which has an interpretation as an open string moduli space \cite{Aganagic:2000gs}. For more general compact geometries, $\tau$ could perhaps also be related to some open string data captured by a lower dimensional geometry along the lines of Ref.~\cite{Alim:2011rp} and references therein.

\subsection*{Acknowledgments}
We would like to thank An Huang, Albrecht Klemm,  Si Li, Hossein Movasati, Dmytro Shklyarov, Cheng-Chiang Tsai, Cumrun Vafa, Xiaoheng Wang and Baosen Wu for comments, discussions and correspondence. This work has been supported by NSF grants PHY-0937443 and DMS-0804454.


\appendix

\section{Modular forms}\label{appendixA}

We summarize the modular objects that appear in this work.
We define (in the literature the choice for $q$ is a matter of convention, in our paper we shall take $q=\exp{2\pi i \tau}$)

\begin{equation}
\vartheta \left[\!\!\! \begin{array}{c}a\\ b\end{array}\!\!\!\right](z,\tau)=\sum_{n\in \mathbbm{Z}}  q^{{1\over 2}(n+a)^2} e^{2\pi i (n+a)(z+b)} \,.
\end{equation}

The following labels are given to the theta functions:
\begin{eqnarray}
&\theta_1(z,\tau)&=\vartheta \left[\!\!\! \begin{array}{c}1/2\\ 1/2\end{array}\!\!\!\right](u,\tau)\,=\sum_{n\in \mathbbm{Z}+{1\over 2}}  (-1)^{n}q^{{1\over 2}n^2}e^{2\pi i n z}\,,\\
&\theta_2(z,\tau)&=\vartheta \left[\!\!\! \begin{array}{c}1/2\\ 0\end{array}\!\!\!\right](u,\tau)\,=\sum_{n\in \mathbbm{Z}+{1\over 2}}  q^{{1\over 2}n^2}e^{2\pi i n z}\,,\\
&\theta_3(z,\tau)&=\vartheta \left[\!\!\! \begin{array}{c}\,\,\,0\,\,\,\\ \,\,\,0\,\,\,\end{array}\!\!\!\right](u,\tau)\,=\sum_{n\in \mathbbm{Z}}  q^{{1\over 2}n^2}e^{2\pi i n z}\,,\\
&\theta_4(z,\tau)&=\vartheta \left[\!\!\! \begin{array}{c}0\\ 1/2\end{array}\!\!\!\right](u,\tau)\,=\sum_{n\in \mathbbm{Z}} (-1)^n q^{{1\over 2}n^2}e^{2\pi in z}\,.
\end{eqnarray}

We also define the following $\theta$--constants:
\begin{equation}
\theta_{2}(\tau)=\theta_2(0,\tau),\quad \theta_{3}(\tau)=\theta_3(0,\tau),\quad \theta_{4}(\tau)=\theta_2(0,\tau)\,.
\end{equation}
The $\eta$--function is defined by
\begin{equation}
\eta(\tau)=q^{\frac{1}{24}}\prod_{n=1}^\infty(1-q^n)\,.
\end{equation}
It transforms according to
\begin{equation}\label{etatrafo}
\eta(\tau+1)=e^{\frac{i\pi}{12}}\eta(\tau),\qquad \eta\left(-\frac{1}{\tau}\right)=\sqrt{\frac{\tau}{i}}\,\eta(\tau)\,.
\end{equation}
The Eisenstein series are defined by
\begin{equation}\label{eisensteinseries}
E_k(\tau)=1-\frac{2k}{B_k}\sum_{n=1}^\infty\frac{n^{k-1}q^n}{1-q^n},
\end{equation}
where $B_k$ denotes the $k$-th Bernoulli number. $E_k$ is a modular form of weight $k$ for $k>2$ and even. The discriminant form and the $j$
invariant are given by
\begin{eqnarray}
   \Delta(\tau) &=& \frac{1}{1728}\left({E_4}(\tau)^3-{E_6}(\tau)^2\right) = \eta(\tau)^{24},\\
    j(\tau)& =& 1728{E_{4}(\tau)^{3}\over E_{4}(\tau)^3-{E_6}(\tau)^2 }\,.
\end{eqnarray}

The following equalities are used a lot throughout our discussions
\begin{eqnarray}
\partial_{\tau}\log \eta(\tau)&=&{1\over 24}E_{2}(\tau)\,,\\
\partial_{\tau}\log \sqrt{\textrm{Im}~\tau}|\eta(\tau)|^2&=&{1\over 24}\widehat{E_{2}}(\tau,\bar{\tau})\, .
\end{eqnarray}
where again by $\partial_{\tau}$ we mean ${1\over 2\pi i}{\partial\over \partial \tau}$.

One can associate to any lattice $\Lambda$ a theta function,
\begin{equation}
in\Theta_{\Lambda}(\tau)=\sum_{x\in \Lambda}e^{2\pi i\cdot  {1\over 2}  ||x||^2\tau},
\end{equation}
see Ref.~\cite{Zagier:2008} and references therein for details on this.

In the following we give the $\theta$--expansions for the
generators of the ring of modular forms and their relations to the
Eisenstein series for the groups $\Gamma_0(N)$, with $N=1^{*},2,3,4$:

\begin{eqnarray}
&N=1^{*}:&\, A(\tau)=\Theta_{E_{8}}^{1\over 4}(\tau)\,.\\
&N=2:&\,
A(\tau)=\Theta_{D_{4}}^{1\over 2}(\tau)\,,\quad
B(\tau)= \theta_{4}^{2}(2\tau)\,,\quad
C(\tau)=2^{-1\over 2}\theta_{2}^{2}(\tau)\,.\\
&N=3:&\\
& A(\tau)&=\sum_{(m,n)\in \mathbb{Z}^{2}}q^{m^{2}-mn+n^2}=\Theta_{A_{2}}(\tau)=
\theta_{2}(2\tau)\theta_{2}(6\tau)+\theta_{3}(2\tau)\theta_{3}(6\tau),\\
&B(\tau)&=\sum_{(m,n)\in \mathbb{Z}^{2}}e^{2\pi i {m-n\over 3}}q^{m^{2}-mn+n^2}\,,\\
&C(\tau)&=\sum_{(m,n)\in \mathbb{Z}^{2}}e^{2\pi i {m-n\over 3}}q^{m^{2}-mn+n^2+m-n}
={1\over 2}(A({\tau\over 3})-A(\tau))\,,\\
&&=\vartheta \left[\!\!\! \begin{array}{c}0\\ 0\end{array}\!\!\!\right](0,\tau)\,
\vartheta \left[\!\!\! \begin{array}{c}1/3\\ 0\end{array}\!\!\!\right](0,3\tau)\,+
\vartheta \left[\!\!\! \begin{array}{c}2/3\\ 0\end{array}\!\!\!\right](0,\tau)\,+
\vartheta \left[\!\!\! \begin{array}{c}1/6\\ 0\end{array}\!\!\!\right](0,3\tau)\,.\\
&N=4:&\, A(\tau)=\Theta_{A_{1}\oplus
A_{1}}(\tau)=\theta_{3}^{2}(2\tau),\quad
B(\tau)=\theta_{4}^{2}(2\tau),\quad C(\tau)=\theta_{2}^{2}(2\tau)\,.
\end{eqnarray}
The generators for the $N=4$ case should be compared to the ring of even weight modular forms with respect
to the principal congruence group $\Gamma(2)$, which is generated by any two of $\theta_{3}^4(\tau),\theta_{2}^4(\tau)
,\theta_{4}^4(\tau)$ since $\theta_{3}^4(\tau)=\theta_{2}^4(\tau)+\theta_{4}^4(\tau)$. Note that the group $\Gamma(2)$
is isomorphic to $\Gamma_{0}(4)$. The are also some nice relations among these generators and the ordinary Eisenstein series $E_{4},E_{6}$:
\begin{eqnarray}
&N=2:& \,
B^{4}+4C^{4}=E_{4}\,,\quad A^{2}(B^4-8C^4)=E_{6}\,.\\
&N=3:& \,
A^{4}+8AC^{3}=E_{4}\,,\quad A^{6}-20A^{3}C^{3}-8C^{6}=E_{6}\,.\\
&N=4:& \,
B^{4}+16B^2C^2+16C^4=E_{4}\,,\quad B^6-30 B^4 C^2-96 B^2 C^4-64 C^6=E_{6}\,.
\end{eqnarray}


\section{Monodromy group and periods of the mirror curve for local $\mathbbm{P}^2$}\label{localp2}
In the following the solutions of $\mathcal{L}_c$ around various points in moduli space will be studied, $z_c$ will henceforth be denoted by $z$.
\subsection{Around $z=0$}
Solutions of the operator $\mathcal{L}_c$ can be found by making a power series ansatz
$$f(z)=\sum_{m=0}^\infty a_m \, z^m\, ,$$
and solving the recursion
\begin{equation}
a_m =\frac{27(m-1/3)(m-2/3)}{m^2} \,a_{m-1}\, ,
\end{equation}
which is solved by
\begin{equation}
\omega_o(z)= \frac{1}{\Gamma(1/3)\Gamma(2/3)}\sum_{m}^\infty \frac{\Gamma(m+1/3)\Gamma(m+2/3)}{\Gamma(m+1)^2} \left(27 z\right)^m\,,
\end{equation}
a second solution can be found by using the Frobenius method:
$$\omega_1(z)= \frac{1}{2\pi i}\frac{d}{d\rho} \left( \sum_{m}^\infty \frac{\Gamma(m+\rho+1/3)\,\Gamma(m+\rho+2/3)}{\Gamma(m+\rho+1)^2} \left(27z\right)^{m+\rho}\right)|_{\rho=0} \,,$$
which gives
\begin{eqnarray}
\omega_1(z)&=&\frac{1}{2\pi i} \omega_0(z) \log(27 z) +  \frac{1}{2\pi i\Gamma(\frac{1}{3})\Gamma(\frac{2}{3})}\sum_{m}^\infty \frac{\Gamma(m+1/3)\Gamma(m+2/3)}{\Gamma(m+1)^2} \left(27z\right)^m \times \nonumber\\
  &&\left( \Psi(m+1/3)+\Psi(m+2/3)-2\Psi(m+1)\right)\, ,
\end{eqnarray}
These solutions can also be given in a compact form using hypergeometric functions:
\begin{eqnarray}
\omega_0(z)= {}_2F_1\left(\frac{1}{3},\frac{2}{3},1,27z\right)\, ,\nonumber\\
\omega_1(z)=\frac{i}{\sqrt{3}}\,{}_2F_1\left(\frac{1}{3},\frac{2}{3},1,1-27z\right)\, ,
\end{eqnarray}
the normalization has been chosen such that $\omega_1= \frac{1}{2\pi i}\omega_0 \log z+ \dots$ and the monodromy becomes
\begin{equation}
M_{z=0}=\left(\begin{array}{cc}
1&0\\
1&1
\end{array} \right)\, .
\end{equation}
The modular parameter of the curve in this region in moduli space is given by $\tau=\frac{\omega_1}{\omega_0}$.

The solutions $\omega_0(z),\omega_1(z)$ can be written as Barnes integrals which will enable us to find the analytic continuation to the $z=\infty$ locus. These expressions are given by, see for example Ref.~\cite{Erdelyi:1981},

\begin{eqnarray}\label{Barnes}
\omega_0(z)=\frac{1}{2\pi i \Gamma(\frac{1}{3})\Gamma(\frac{2}{3})} \int_C \frac{\Gamma(-s)\Gamma(s+\frac{1}{3})\Gamma(s+\frac{2}{3})}{\Gamma(s+1)} (-27z)^s\,,\quad |\textrm{arg}~(-z)|<\pi,\\
\omega_1(z)=\frac{1}{4\pi^2 \Gamma(\frac{1}{3})\Gamma(\frac{2}{3})} \int_C \Gamma(-s)^2 \Gamma(s+\frac{1}{3})\Gamma(s+\frac{2}{3}) (27z)^s\,,\quad |\textrm{arg}~(-z)|<\pi.
\end{eqnarray}

\subsection{Around $z =1/27$}
Choosing a coordinate $y=1/27-z$ we can compute the monodromies around this expansion locus.  The periods become
\begin{eqnarray}
\omega_0(y)={}_2F_1(\frac{1}{3},\frac{2}{3},1,1- 27 y)\, ,\nonumber\\
\omega_1(y)=\frac{i}{\sqrt{3}}\,{}_2F_1(\frac{1}{3},\frac{2}{3},1,27 y)\, ,
\end{eqnarray}
with monodromy
\begin{equation}
M_{z=1/27}=\left(\begin{array}{cc}
1&-3\\
0&1
\end{array} \right)\, .
\end{equation}
The modular parameter in this case is given by
\begin{equation}
\tau_D=-\frac{1}{3}\frac{\omega_0}{\omega_1}=-\frac{1}{3\tau}\,.
\end{equation}

The transformation from $\tau$ to $\tau_D$ is
\begin{equation}
W_{N}=\left(\begin{array}{cc}
0&-1\\
3&0
\end{array} \right) \,.
\end{equation}

\subsection{Around $z=\infty$}
\subsubsection{Solving in local coordinates}
Using the local coordinate $x=z^{-1}$ the operator $\mathcal{L}_c$ becomes
\begin{equation}
\mathcal{L}_c=x\,\theta_x^2-27(\theta_x-1/3)(\theta_x-2/3)\, ,
\end{equation}
acting with this operator on the ansatz $f(x)=\sum_{m=0}^\infty a_m z^{m+p}$, where $p\in \mathbbm{Q}$ is determined by solving the indicial equation. We find $p=1/3,2/3$. We find two solutions:
\begin{eqnarray}
f_0(x)=x^{1/3}\frac{\Gamma(2/3)}{\Gamma(1/3)^2} \sum_{m=0}^\infty \frac{\Gamma(m+1/3)^2}{\Gamma(m+1)\Gamma(m+2/3)} \left( \frac{x}{27}\right)^m\, , \nonumber\\
f_1(x)= x^{2/3}\frac{\Gamma(4/3)}{\Gamma(2/3)^2} \sum_{m=0}^\infty \frac{\Gamma(m+2/3)^2}{\Gamma(m+1)\Gamma(m+4/3)} \left( \frac{x}{27}\right)^m\, .
\end{eqnarray}
These solutions diagonalize the monodromy around $z=\infty$. As $x\rightarrow e^{2\pi i}x$, the solutions transform according to:
\begin{equation}
\left(\begin{array}{c}
f_0(x)\\
f_1(x)
\end{array} \right)
\rightarrow \left(\begin{array}{cc}
e^{2\pi i/3}&0\\
0&e^{4\pi i/3}
\end{array} \right)\, \left(\begin{array}{c}
f_0(x)\\
f_1(x)
\end{array} \right)\,.
\end{equation}

\subsubsection{Analytic continuation}
In the following we analytically continue $\omega_0(z),\omega_1(z)$ and express these in the basis $f_0(x),f_1(x)$. Using the expressions in terms of Barnes integrals in Eq.(\ref{Barnes}) and closing the contour on the left we find ($\alpha=e^{-i\pi/3}$):
\begin{eqnarray}
\omega_0(x)&=&\frac{\alpha}{3} \frac{\Gamma(1/3)}{\Gamma(2/3)^2} f_0(x) -\frac{\alpha^2}{3} \frac{\Gamma(2/3)}{\Gamma(1/3)^2} f_1(x)\, ,\nonumber\\
\omega_1(x)&=&\frac{i}{3\sqrt{3}} \frac{\Gamma(1/3)}{\Gamma(2/3)^2} f_0(x) -\frac{i}{3\sqrt{3}} \frac{\Gamma(2/3)}{\Gamma(1/3)^2} f_1(x)\,.
\end{eqnarray}
Knowing the monodromy of the solutions $f_0(x),f_1(x)$ we can now compute:
\begin{equation}
\left(\begin{array}{c}
\omega_0(x)\\
\omega_1(x)
\end{array} \right)
\rightarrow \left(\begin{array}{cc}
1&3\\
-1&-2
\end{array} \right)\, \left(\begin{array}{c}
\omega_0(x)\\
\omega_1(x)
\end{array} \right) \, .
\end{equation}
We verify:
\begin{equation}
\left(\begin{array}{cc}
1&3\\
-1&-2
\end{array} \right)^{-1}\, = \left(\begin{array}{cc}
1&-3\\
0&1
\end{array} \right)\,  \cdot \left(\begin{array}{cc}
1&0\\
1&1
\end{array} \right)\,  .
\end{equation}
Hence the monodromy subgroup of $SL(2,\mathbbm{Z})$ is $\Gamma_0(3)$.

\section{Higher genus amplitudes}\label{highergenus}
\subsection{Local $\mathbbm{P}^2$}
\begin{eqnarray}
F^{(3)}&=&\frac{5359 C_0^4+12572 C_0^3+9722 C_0^2+2668 C_0+157}{8709120}\nonumber\\
&&-  \frac{ \left(211 C_0^4+557 C_0^3+500
   C_0^2+169 C_0+15\right) G_1^{10} T_2 } {  51840G_{1}^{12}}\nonumber\\
&&+\frac{ 2 (C_0+1)^2 \left(293
   C_0^2+277 C_0+45\right) G_1^8 T_2^2    -\frac{5}{3} (C_0+1)^2 \left(529
   C_0^2+678 C_0+173\right) G_1^6 T_2^3} {  51840G_{1}^{12}}\nonumber\\
&& + \frac{40 (C_0+1)^3 (20 C_0+13) G_1^4
   T_2^4-500 (C_0+1)^4 G_1^2 T_2^5+200 (C_0+1)^4
   T_2^6} {51840G_{1}^{12}}\,. \\
  F^{(3)}_D(t_c)&=& \frac{1}{59521392 t_c^4}-\frac{1}{117573120}-\frac{t_c}{54432}+\frac{23855
   t_c^2}{246903552}-\frac{557 t_c^3}{1259712}\nonumber\\
  &&+\frac{15575867 t_c^4}{8465264640}+\mathcal{O}(t_c^5)
\end{eqnarray}

\subsection{Local $E_8$ del Pezzo}
\label{sec:local-e_8-del}

\begin{equation}
  \label{eq:1}
  \begin{aligned}
    F^{(3)} &= {\frac {5}{16}}\,{\frac { \left( 1+C_{{0}} \right) ^{4}{T_{{2}}}^{6}}{
{G_{{1}}}^{12}}}-{\frac {5}{12}}\,{\frac { \left( 1+C_{{0}} \right) ^{
4}{T_{{2}}}^{5}}{{G_{{1}}}^{10}}}\\
   &\phantom{=}+{\frac {1}{768}}\,{\frac { \left( 1+
C_{{0}} \right) ^{2} \left( 175+362\,C_{{0}}+215\,{C_{{0}}}^{2}
 \right) {T_{{2}}}^{4}}{{G_{{1}}}^{8}}}\\
   &\phantom{=}-{\frac {1}{10368}}\,{\frac {
 \left( 1+C_{{0}} \right)  \left( 1865\,C_{{0}}+2743\,{C_{{0}}}^{2}+
1234\,{C_{{0}}}^{3}+650 \right) {T_{{2}}}^{3}}{{G_{{1}}}^{6}}}\\
   &\phantom{=}+{\frac
{1}{1658880}}\,{\frac { \left( 13875+20716\,C_{{0}}+115394\,{C_{{0}}}^
{2}+52387\,{C_{{0}}}^{4}+138540\,{C_{{0}}}^{3} \right) {T_{{2}}}^{2}}{
{G_{{1}}}^{4}}}\\
   &\phantom{=}-{\frac {1}{2488320}}\,{\frac { \left( 1+C_{{0}}
 \right)  \left( 1263\,C_{{0}}+11997\,{C_{{0}}}^{2}+11684\,{C_{{0}}}^{
3}+1300 \right) T_{{2}}}{{G_{{1}}}^{2}}}\\
   &\phantom{=}
    -{\frac {365}{41472}}+{\frac {1}{5760}}\,\chi+{\frac {413}{89579520}}
\,C_{{0}}+{\frac {436981}{2508226560}}\,{C_{{0}}}^{2}+{\frac {265373}{
627056640}}\,{C_{{0}}}^{3}\\
   &\phantom{=}+{\frac {1491431}{5016453120}}\,{C_{{0}}}^{4}\\
    F_D^{(3)}(t_c) &= {\frac {1}{35107145515008}}\,{t_{{c}}}^{-4}+{\frac {528257}{4232632320
}}-{\frac {1}{1451520}}\,\chi-{\frac {9393421}{334430208}}\,t_{{c}}\\
   &\phantom{=}+{
\frac {645246474275}{142216445952}}\,{t_{{c}}}^{2}-{\frac {
24963215980267}{40633270272}}\,{t_{{c}}}^{3}+{\frac {11259598289900599
}{152374763520}}\,{t_{{c}}}^{4}\\
   &\phantom{=}+O \left( {t_{{c}}}^{5} \right)
  \end{aligned}
\end{equation}

\subsection{Local $E_7$ del Pezzo}
\label{sec:local-e_7-del}

\begin{equation}
  \label{eq:2}
  \begin{aligned}
    F^{(3)} &= 5\,{\frac { \left( 1+C_{{0}} \right) ^{4}{T_{{2}}}^{6}}{{G_{{1}}}^{12}
}}-{\frac {5}{24}}\,{\frac { \left( 1+C_{{0}} \right) ^{3} \left( 17+13\,C_{{0}} \right) {T_{{2}}}^{5
}}{{G_{{1}}}^{10}}}\\
   &\phantom{=}+{\frac {1}{2304}}\,
{\frac { \left( 1+C_{{0}} \right) ^{2} \left( 2387+3582\,
C_{{0}}+1855\,{C_{{0}}}^{2} \right) {T_{{2}}}^{4}}{{G_{{1}}}^{8}}}\\
   &\phantom{=}-{\frac {1}{82944
}}\,{\frac { \left( 1+C_{{0}} \right) \left( 13057+23151
\,C_{{0}}+31323\,{C_{{0}}}^{2}+13129\,{C_{{0}}}^{3} \right) {T_{{2}}}^{3}}{{G_{{1}}
}^{6}}}\\
   &\phantom{=}+{\frac {1}{8847360}}\,{\frac { \left( 112685+
17636\,C_{{0}}+388806\,{C_{{0}}}^{2}+460484\,{C_{{0}}}^{3}+171029\,{C_
{{0}}}^{4} \right) {T_{{2}}}^{2}}{{G_{{1}}}^{4}}}\\
   &\phantom{=}-{\frac {1}{70778880}}\,{\frac {
 \left( 1+C_{{0}} \right) \left( 36625+20751\,C_{{0}}+92651\,{
C_{{0}}}^{2}+92325\,{C_{{0}}}^{3} \right) T_{{2}} }{{G_{{1}}}^{2}}}\\
   &\phantom{=}
   +{\frac {11269}{1358954496}}-{\frac {14473}{1698693120}}\,C_{{0}}+{
\frac {25157}{880803840}}\,{C_{{0}}}^{2}+{\frac {607447}{11890851840}}
\,{C_{{0}}}^{3}\\
   &\phantom{=}+{\frac {1767811}{47563407360}}\,{C_{{0}}}^{4}-{\frac {
1}{1451520}}\,\chi\\
    F_D^{(3)}(t_c) &={\frac {1}{67645734912}}\,{t_{{c}}}^{-4}+{\frac {289321}{2972712960}}-
{\frac {1}{1451520}}\,\chi-{\frac {21932749}{4227858432}}\,t_{{c}}\\
   &\phantom{=}+{
\frac {37014805643}{199766310912}}\,{t_{{c}}}^{2}-{\frac {11337784009}
{2113929216}}\,{t_{{c}}}^{3}+{\frac {9696390763877}{71345111040}}\,{t_
{{c}}}^{4}+O \left( {t_{{c}}}^{5} \right)
  \end{aligned}
\end{equation}

\subsection{Local $E_6$ del Pezzo}
\label{sec:local-e_6-del}

\begin{equation}
  \label{eq:6}
  \begin{aligned}
    F^{(3)} &= {\frac {405}{16}}\,{\frac { \left( 1+C_{{0}} \right) ^{4}{T_{{2}}}^{6}
}{{G_{{1}}}^{12}}}-{\frac {45}{32}}\,{\frac { \left( 1+C_{{0}}
 \right) ^{3} \left( 9+5\,C_{{0}} \right) {T_{{2}}}^{5} }{{G_{{1}}}^{10
}}}\\
   &\phantom{=}+\frac{1}{8}\,{\frac { \left( 1+C_{{0}} \right) ^{2}\left( 21
+22\,C_{{0}}+10\,{C_{{0}}}^{2} \right) {T_{{2}}}^{4}  }{{G_{{1}}}^{8}}}\\
   &\phantom{=}-{\frac {1}{
3456}}\,{\frac { \left( 1+C_{{0}} \right) \left( 993+
1083\,C_{{0}}+1387\,{C_{{0}}}^{2}+529\,{C_{{0}}}^{3} \right) {T_{{2}}}^{3} }{{G_{{1}
}}^{6}}}\\
   &\phantom{=}+{\frac {1}{25920}}\,{\frac {\left( 450-88\,C_{{0
}}+807\,{C_{{0}}}^{2}+838\,{C_{{0}}}^{3}+293\,{C_{{0}}}^{4} \right) {T_{{2}}}^{2} }{
{G_{{1}}}^{4}}}\\
   &\phantom{=}-{\frac {1}{466560}}\,{\frac { \left( 1+C_{{0}}
 \right) \left( 255+142\,C_{{0}}+258\,{C_{{0}}}^{2}+211\,{C_{{0
}}}^{3} \right) T_{{2}} }{{G_{{1}}}^{2}}}\\
   &\phantom{=}+{\frac {47}{6718464}}-{\frac {323}{35271936}}\,C_{{0}}+{\frac {1219}{
117573120}}\,{C_{{0}}}^{2}+{\frac {323}{25194240}}\,{C_{{0}}}^{3}\\
   &\phantom{=}+{
\frac {5359}{705438720}}\,{C_{{0}}}^{4}-{\frac {1}{1451520}}\,\chi\\
    F_D^{(3)}(t_c) &={\frac {1}{4821232752}}\,{t_{{c}}}^{-4}+{\frac {443641}{4761711360}}-{
\frac {1}{1451520}}\,\chi-{\frac {12769}{4408992}}\,t_{{c}}\\
   &\phantom{=}+{\frac {
1128955631}{19999187712}}\,{t_{{c}}}^{2}-{\frac {88674605}{102036672}}
\,{t_{{c}}}^{3}+{\frac {7946436569147}{685686435840}}\,{t_{{c}}}^{4}+O
 \left( {t_{{c}}}^{5} \right)
  \end{aligned}
\end{equation}

\subsection{Local $E_5$ del Pezzo}
\label{sec:local-e_5-del}

\begin{equation}
  \label{eq:9}
  \begin{aligned}
    F^{(3)} &= 80\,{\frac { \left( 1+C_{{0}} \right) ^{4}{T_{{2}}}^{6}}{{G_{{1}}}^{12
}}}-\frac{10}{3}\,{\frac { \left( 1+C_{{0}} \right) ^{3} \left(
11+2\,C_{{0}} \right) {T_{{2}}}^{5} }{{G_{{1}}}^{10}}}\\
   &\phantom{=}+{\frac {1}{144}}\,{\frac {
 \left( 1+C_{{0}} \right) ^{2} \left( 1007+312\,C_{{0}}+
145\,{C_{{0}}}^{2} \right) {T_{{2}}}^{4}}{{G_{{1}}}^{8}}}\\
   &\phantom{=}-{\frac {1}{5184}}\,{
\frac { \left( 1+C_{{0}} \right) \left( 3679+690\,C_{{0}
}+1605\,{C_{{0}}}^{2}+398\,{C_{{0}}}^{3} \right) {T_{{2}}}^{3}}{{G_{{1}}}^{6}}}\\
   &\phantom{=}+{
\frac {1}{552960}}\,{\frac { \left( 22325-10912\,C_{{0}}+
15962\,{C_{{0}}}^{2}+6552\,{C_{{0}}}^{3}+1689\,{C_{{0}}}^{4} \right) {T_{{2}}}^{2} }
{{G_{{1}}}^{4}}}\\
   &\phantom{=}-{\frac {1}{4423680}}\,{\frac { \left( 5395+
1586\,C_{{0}}+9926\,{C_{{0}}}^{2}+3228\,{C_{{0}}}^{3}+1647\,{C_{{0}}}^
{4}+298\,{C_{{0}}}^{5} \right) T_{{2}}}{{G_{{1}}}^{2} \left( 1+C_{{0}}
 \right) }}\\
   &\phantom{=}
  +{\frac {1297}{84934656}}-{\frac {3569}{212336640}}\,C_{{0}}+{\frac {
1151}{247726080}}\,{C_{{0}}}^{2}+{\frac {3847}{1486356480}}\,{C_{{0}}}
^{3}\\
   &\phantom{=}+{\frac {373}{594542592}}\,{C_{{0}}}^{4}-{\frac {1}{1451520}}\,
\chi-{\frac {1}{6193152}}\,{\frac {C_{{0}} \left( 457+220\,C_{{0}}
 \right) }{ \left( 1+C_{{0}} \right) ^{2}}}|\\
   F_D^{(3)}(t_c) &= {\frac {1}{1056964608}}\,{t_{{c}}}^{-4}-{\frac {1}{1451520}}\,\chi+{
\frac {1429}{46448640}}-{\frac {229921}{198180864}}\,t_{{c}}\\
   &\phantom{=}+{\frac {
62419963}{3121348608}}\,{t_{{c}}}^{2}-{\frac {72381647}{297271296}}\,{
t_{{c}}}^{3}+{\frac {2724837497}{1114767360}}\,{t_{{c}}}^{4}+O \left(
{t_{{c}}}^{5} \right)
  \end{aligned}
\end{equation}


\end{document}